\documentclass[journal]{IEEEtran}
\ifCLASSINFOpdf

\else

\fi
\usepackage[cmex10]{amsmath}
\usepackage{amssymb}
\usepackage{cite}
\usepackage{graphicx}
\usepackage{array,color}
\usepackage{amsmath}
\usepackage{stfloats}
\usepackage{multirow}
\usepackage{graphicx}
\usepackage{subfigure}
\usepackage{tabularx}
\usepackage{epsfig,epsf,color,balance,cite}
\usepackage{algorithmic}
\usepackage{algorithm}
\usepackage{bm}
\usepackage{textcomp}

\begin{document}

\title{Power-and Rate-Adaptation Improves the Effective Capacity of C-RAN for Nakagami-$m$ Fading Channels}

\author{
Hong Ren, Nan Liu, Cunhua Pan, Maged Elkashlan, Arumugam Nallanathan, \IEEEmembership{Fellow, IEEE}, Xiaohu You, \IEEEmembership{Fellow, IEEE} and Lajos Hanzo, \IEEEmembership{Fellow, IEEE}
\thanks{H. Ren, N. Liu and X. You are with the Southeast University, Nanjing, China. (e-mail:\{renhong, nanliu, xhyu\}@seu.edu.cn). C. Pan, M.Elkashlan and Arumugam Nallanathan are with the Queen Mary University of London, London E1 4NS, U.K. (Email:\{c.pan, maged.elkashlan, a.nallanathan\}@qmul.ac.uk). L. Hanzo is with the School of Electronics and Computer Science, University of Southampton, Southampton, SO17 1BJ, U.K. (e-mail:lh@ecs.soton.ac.uk). }

\thanks{This work is partially supported by the National Natural Science Foundation of China under Grants $61571123$ and $61521061$, the National 863 Program (2014AA01A702). L.Hanzo would like to acknowledge the financial support of the European Research  Council, Advanced Fellow grant Beam-Me-Up.}

}

\maketitle

\begin{abstract}
We propose a  power-and rate-adaptation scheme for cloud radio access networks (C-RANs), where each radio remote head (RRH) is connected to the baseband unit (BBU) pool through  optical links. The RRHs jointly support the users by efficiently exploiting the enhanced spatial degrees of freedom. Our proposed scheme aims for maximizing the effective capacity (EC) of the user subject to both per-RRH average-and peak-power constraints, where the EC is defined as the maximum arrival rate that can be supported by the C-RAN under the statistical delay requirement. We first transform the EC maximization problem into an equivalent convex optimization problem. By using the Lagrange dual decomposition method and solving the Karush-Kuhn-Tucker (KKT) equations, the optimal transmission power of each RRH can be obtained in closed-form. Furthermore, an online tracking method is provided for approximating the average power of each RRH. For the special case of two RRHs, the expression of the average power of each RRH can be calculated in explicit form. Hence, the Lagrange dual variables can be computed in advance in this special case. Furthermore, we derive the power allocation for two important extreme cases: 1) no delay constraint; 2) extremely stringent delay-requirements. Our simulation results show that the proposed scheme significantly outperforms the conventional algorithm without considering the delay requirements. Furthermore, when appropriately tuning the value of the delay exponent, our proposed algorithm is capable of guaranteeing a delay outage probability below $10^{-9}$ when the maximum tolerable delay is 1 ms. This is suitable for the future ultra-reliable low latency communications (URLLC).
\end{abstract}


%
\section{Introduction}

The fifth-generation (5G) wireless system to be deployed by 2020 is expected to offer a substantially increased  capacity \cite{Andrews14}. To achieve this ambitious goal, the C-RAN concept has been regarded as one of the most promising  solutions. In particular, C-RAN is composed of three key components: 1) a pool of BBUs centrally located at a cloud data center; 2) low-cost, low-power distributed RRHs deployed in the network; 3) high-bandwidth low latency fronthaul links that connect the RRHs to the BBU pool. Under the C-RAN architecture, most of the baseband signal processing of conventional base stations has been shifted to the BBU pool and the RRHs are only responsible for simple transmission/reception functions,  some of the hitherto centralized signal processing operations can be relegated to the BBU pool. Hence, the network capacity  can be significantly improved.

Recently, the performance of C-RAN has been extensively studied \cite{mugenpengwcl,shiyuan2014,binbindaijsac,pan2017joint}, albeit these papers have been focused on the physical layer issues, which giving no cognizance to the delay of the upper layer's. However, most of the multimedia services, such as video conferencing and mobile TV have stringent delay requirements. Due to the time-varying characteristics of fading channels, it is impossible to impose a deterministic delay-bound guarantee for wireless communications. For the sake of analyzing the statistical QoS performance, Wu \emph{et al}. \cite{Dapeng2003} introduced the notion of effective capacity (EC), which can be interpreted as the maximum constant packet arrival rate that can be supported by the system, whilst satisfying a maximum buffer-violation probability constraint.

Due to the complex expression of EC, most of the existing papers focus on the EC maximization problem for the simple scenario, where there is only a single transmitter \cite{Tang2007,Musavian2010,xizhang2013,Yli15,cheng2016}. Specifically, a QoS-driven
power-and rate-adaptation scheme was proposed for single-input single-output
(SISO) systems communicating over flat-fading channels in \cite{Tang2007}, with the objective of maximizing EC subject to both delay-QoS and average power constraints, which is characterized by the QoS exponent $\theta$. A smaller $\theta$ corresponds to a looser QoS guarantee, while a higher value of $\theta$ represents a more stringent QoS requirement. The results of \cite{Tang2007} showed that in the extreme case of  $\theta\rightarrow 0$, the power allocation reduces to the conventional water-filling solution. By contrast, when $\theta\rightarrow \infty$, the optimal power allocation becomes the channel inversion scheme, where the system operates under a fixed transmission rate. The EC maximization problem was studied in \cite{Musavian2010} in the context of cognitive radio networks, where the power constraints of \cite{Tang2007} were replaced by the maximum tolerable interference-power at the primary user. Closed-form expressions of both the power allocation and  the EC were derived for the secondary user.  The power minimization problem subject to EC constraints was considered in \cite{xizhang2013} for three different scenarios. Significant power savings can be achieved by using the power allocation scheme of \cite{xizhang2013}. To a further advance, both subcarrier and power allocation were investigated in \cite{Yli15} for a one-way relay network wherein the optimal subcarrier and power allocation was derived by adopting the Lagrangian dual decomposition method. Most recently, Wenchi \emph{et al}. \cite{cheng2016} considered both the average-and peak-power constraints when maximizing the EC, and provided the specific conditions, when the peak power constraints can be removed. From the information-theoretical point of view, it is difficult to solve the ergodic capacity maximization problem subject to both the average-and peak-power constraints, even for the basic Gaussian white noise channel. In Shannon's landmark paper \cite{shannon2001mathematical}, only the asymptotically low and high SNR for the bandlimited continuous time Gaussian channel was studied. This problem was later studied by Smith \cite{smith1971information} and showed that the capacity achieving distribution is discrete. In \cite{Shamai95}, the authors considered the more general case of the quadrature additive Gaussian channel and derived the lower and upper bounds of this channel's capacity. Khojastepour \emph{et al.} \cite{Khojastepour04} studied the capacity of the fading channel under both types of power constraints under the assumption of the perfect channel state information available at both the transmitter and receiver side. However, the above papers \cite{shannon2001mathematical,smith1971information,Shamai95} neither consider the upper layer delay nor the multiple transmission points.

To avoid any traffic congestion and attain a good C-RAN performance, the adaptive power allocation scheme should take the diverse QoS requirements into account to guarantee the satisfaction of users. In this paper, we aim for jointly optimizing the power allocation of each RRH in order to maximize the EC of a user of the C-RAN, where both the average power and peak power constraints of each RRH are considered. This user is jointly served by all RRHs of the C-RAN due to the powerful computational capability of the BBU pool. Unfortunately, the power allocation schemes developed in the aforementioned papers for single-transmitter scenarios     \cite{Tang2007,Musavian2010,xizhang2013,Yli15,cheng2016} cannot be directly
applied to C-RAN's relying on multiple RRHs for serving the user and to simultaneously exploit the spatial degrees of freedom. The reason can be explained as follows. In these papers, there is only a single transmitter and only a sum-power constraint is imposed. The Lagrange method can be used to find the optimal power allocation, which is in the form of a water-filling-like solution in general. However, for the C-RAN, all RRHs have their individual power constraints and the power cannot be shared among the RRHs. Yu \emph{et al.} \cite{QYYUTCOM} provided a detailed reason as to why the classic Lagrange method cannot be readily applied in C-RAN. Hence, new methods have to be developed. Specifically, the contributions of this paper can be summarized as follows:
\begin{enumerate}
  \item In this paper, we derive the optimal power allocation for each RRH by resorting to the Lagrangian dual decomposition and the Karush-Kuhn-Tucker (KKT) conditions. The power allocation solutions depend both on the user's QoS requirements and on the joint channel conditions of the RRHs.
  \item The Lagrangian dual decomposition requires us to calculate the subgradient, where the average power of each iteration should be obtained. However, it is numerically challenging to derive the expression of average power for each RRH. To tackle this issue, we provide an online training method for tracking the average power of each RRH. For the special case of a single RRH, the closed-form expression of average power can be obtained. For the more complex case of two RRHs, we also provide the expression of the average power for each RRH in explicit form, which can be numerically evaluated.
 \item We provide the closed-form power allocation solutions for two extremely important cases: 1) When the QoS exponent $\theta$ approaches zero, which corresponds to the conventional ergodic capacity maximization problem; 2) When the QoS exponent $\theta$ tends to infinity, which represents the strictly delay-limited case. We also extend our work to the more general multiuser case, where the optimal transmit power for each user is derived.
  \item Our simulation results will show that the proposed algorithms significantly outperforms the existing algorithms. Additionally, by appropriately choosing the QoS exponent, our proposed delay-aware algorithm can guarantee a delay-outage probability below $10^{-9}$, when the maximum tolerable end-to-end (E2E) delay is 1 ms, which satisfies the strict delay requirements of the future ultra-reliable low latency communications (URLLC).
\end{enumerate}

The most closely related paper of our work is \cite{cheng2016}, where single transmitter is considered. The optimal power allocation can be readily derived in closed form by the aid of the Lagrangian dual decomposition, which is in the water-filling-like form. However, our considered C-RAN involves multiple independent transmitters, the power allocation derivations are much more involved, and the power allocation solution is not the same as the conventional water-filling form. In  \cite{cheng2016}, only one dual variable needs to be optimized, which can be obtained by using the bisection search method, while multiple dual variables are involved in the C-RAN scenario and the subgradient method is adopted to update the dual variables. Furthermore, in both papers, the average power expression is required in the updating of dual variables. However, this issue is not studied in \cite{cheng2016}, while our paper provides more details about the analysis of this issue.

\vspace{-0.2cm}
\begin{table}
\renewcommand{\arraystretch}{1.1}
\caption{ The List of Notations}
\label{tabone}
\centering
\begin{tabular}{|c|c|}
\hline
$I$  & The number of RRHs   \\
\hline
$\cal I$ & The set of RRHs \\
\hline
$T_f$ & The time frame length \\
\hline
$m$  & Nakagami parameter\\
\hline
$\alpha_i$ & Channel-power-to-noise ratio (CPNR)\\
\hline
$\theta$ & QoS exponent\\
\hline
${\rm{EC}}\left( \theta  \right)$ & The effective capacity specified by $\theta$\\
\hline
$\mu$  & Constant arrival data rate\\
\hline
${D_{\max }}$ & Maximum delay bound\\
\hline
$B$ & System bandwidth\\
\hline
$P_i^{{\rm{avg}}}$ & Average power constraint of RRH $i$\\
\hline
$P_i^{{\rm{peak}}}$ & Peak power constraint of RRH $i$\\
\hline
 \end{tabular}\vspace{-0.1cm}
\end{table}
The rest of this paper is organized as follows. In Section \ref{systemmodel}, the C-RAN system is introduced along with the concept of EC and our problem formulation. In Section \ref{genralcase}, we provide the optimal power allocation for the general case of any number of RRHs. In Section \ref{specialcaseoftwo}, we derive the integral expressions's closed-form solution concerning the average transmit power of each RRH for the sake of updating the Lagrangian dual variables. In Section \ref{extremeajeoioif}, we obtain the power allocation for two extreme cases, namely for the delay-tolerant scenario and for extremely stringent delay constraints. Numerical results are also provided  in Section \ref{simulation}. Finally, our conclusions are drawn in Section \ref{conclusion}. The other notations are summarized in Table.~\ref{tabone}.

\vspace{-0.2cm}
\section{system model}\label{systemmodel}
\begin{figure}
\centering
\includegraphics[width=1.9in]{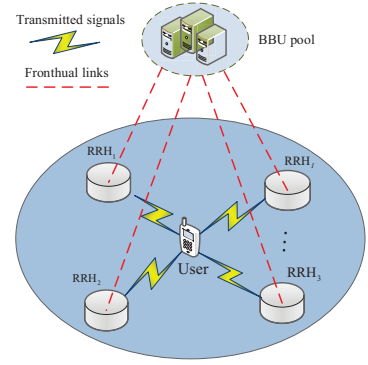}
\caption{Topology of a C-RAN with  $I$ RRHs. }\vspace{-0.2cm}
\label{systemmodel-C-RAN}
\end{figure}

\begin{figure}
\centering
\includegraphics[width=2.6in]{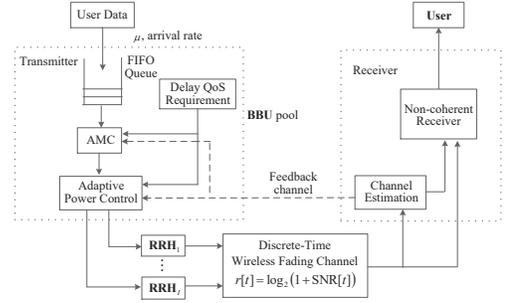}
\caption{Cross-layer transmission model.}\vspace{-0.2cm}
\label{systemmodel-queue}
\end{figure}

Consider a downlink C-RAN consisting of $I$ RRHs and a single user \footnote{The method developed in this paper can also be applied to the multi-user scenario, where all users apply the classical orthogonal frequency division multiplexing access (OFDMA) technique to remove the multi-user interference.}, where each RRH and the user have a single antenna, as depicted in Fig. \ref{systemmodel-C-RAN}. The set of RRHs is denoted as $\mathcal{I}=\{1,2,\cdots, I\}$.  All the RRHs are assumed to be connected to the BBU pool through the fronthaul links relying on high-speed fiber-optic cables.  In Fig.~{\ref{systemmodel-queue}}, the upper layer packets are first buffered in first-in-first-out (FIFO) queue, which will be transmitted to the physical layer of the RRHs. Since the RRHs are only responsible for simple transmission/reception, we do not consider the storage function at the RRHs. At the data-link layer, the upper-layer packets are partitioned into frames and then each frame will be mapped into bit-streams at the physical layer. The channel is assumed to obey the stationary block fading model, implying that they are fixed during each time frame of length of $T_f$, while it is switched independently over different time frames.

To elaborate, we consider a Nakagami-$m$ block-fading channel, which is very general and includes most of the practical wireless communication channels as special cases\cite{simon2005digital}. The parameter $m$ represents the severeness of the channel, where the fading channel fluctuations are reduced with $m$. The probability density function (PDF) of the Nakagami-$m$ channel spanning from the $i$-th RRH to the user is given by:
\begin{equation}\label{rheisghu}
f(\alpha_i)=\frac{\alpha_i^{m-1}}{\Gamma(m)}\left(\frac{m}{\bar{\alpha_i}}\right)^m {\text{exp}}\left(-\frac{m}{\bar\alpha_i}\alpha_i\right), \;\;\;\;\alpha_i\ge 0,
\end{equation}
where $\Gamma(m)=\int_0^\infty w^{m-1}e^{-w}dw$ is the Gamma function, $\alpha_i$ denotes the instantaneous channel-power-to-noise ratio (CPNR) from the $i$th RRH to the user, and $\bar{\alpha_i}$ is the average received CPNR at the user from the $i$th RRH, denoted as ${{P{L_i}} \mathord{\left/
 {\vphantom {{P{L_i}} {{\sigma ^2}}}} \right.
 \kern-\nulldelimiterspace} {{\sigma ^2}}}$, where $PL_i$ is the large-scale fading channel gain spanning from the $i$th RRH to the user that includes the path loss and shadowing effect, and $\sigma^2$ is the noise power.

Let us define $\bm\alpha=\left[ {{\alpha _1},{\alpha _2}, \cdots ,{\alpha _I}} \right]^T$. Since $\alpha_1,\cdots,\alpha_I$ are independent, the joint PDF of $\bm\alpha$ is given by
\begin{equation}
f(\bm\alpha)=f(\alpha_1)f(\alpha_2)\cdots f(\alpha_I).
\label{pdfex}
\end{equation}

\vspace{-0.2cm}

\vspace{-0.2cm}

\vspace{-0.7cm}
\subsection{Effective Capacity (EC)}

This paper considers the E2E delay requirement for each packet. In C-RANs, the E2E delay, denoted as ${D_{\max }}$, includes the uplink (UL) and downlink (DL) transmission delays of ${D_T}$, queueing delay at the buffer of the BBU pool  of  ${D_q}$, and the fronthaul delay of  ${D_F}$. Since the fronthaul links are usually deployed with high-speed fiber, the fronthaul delay is much less than 1 ms \cite{ganzhang2015}. In addition, since the packet size in URLLC is very short, the UL and DL transmission can be finished within a very short time. Hence, the E2E delay is mainly dominated by the queueing delay at the buffer of the BBU pool, which is given by  ${D_q}={D_{\max }}-{D_F}-{D_T}$. In the following, we mainly focus on the study of the queueing delay, which is characterized by the EC.

The EC is defined as the maximum constant frame arrival rate that a given service process can support, while obeying the delay requirement indicated by the QoS exponent $\theta$ that will be detailed later. Let the sequence ${R[k],k=1,2,\cdots}$ represent the data service-rate, which follows a discrete-time stationary and ergodic stochastic process. The parameter $k$ is the time frame index. Let us denote by $S(t)\triangleq \sum\nolimits_{k=1}^{t}R[k]$  the partial sum of the service process over the time sequence spanning from $k=1$ to $k=t$. Let us furthermore assume that the Gartner-Ellis limit of $S[t]$, which is denoted by $\Lambda_C(\theta)=\lim_{t\rightarrow\infty}(1/t)\log(\mathbb{E}\{e^{\theta S[t]}\})$, is a convex differentiable function for
all real-value of $\theta$ \cite{Tang2007}. Then, the EC of the service process specified by $\theta$ is
\begin{equation}
 \text{EC}(\theta)=-\frac{\Lambda_C(-\theta)}{\theta}=-\frac{1}{\theta}\log(\mathbb{E}\{e^{-\theta R[k]}\})
\end{equation}
where $\mathbb{E}\{\cdot\}$ denotes the expectation operator.

Let us assume that a data source enters a queue of infinite buffer size at a constant data rate $\mu$. The probability that the delay exceeds a maximum delay bound of $D_{q}$ satisfies \cite{Dapeng2003}
\begin{equation}\label{frehiutgi}
 P_{{\rm{delay}}}^{{\rm{out}}} = \Pr \left\{ {{\rm{Delay}} \ge {D_{q}}} \right\} \approx \varepsilon {e^{ - \theta \mu {D_{q }}}},
\end{equation}
where $f(x) \approx g(x)$ indicates that $\mathop {\lim }\limits_{x \to \infty } \left[ {{{f(x)} \mathord{\left/
 {\vphantom {{f(x)} {g(x)}}} \right.
 \kern-\nulldelimiterspace} {g(x)}}} \right] = 1$, and $\varepsilon$ is the probability that the buffer is non-empty, which can be calculated by using the method in \cite{Dapeng2003}. The parameter $\theta$ can be found by letting the constant arrival rate to be equal to the EC, i.e. setting it to $\mu  = {\rm{EC}}(\theta )$. Hence, if the delay-bound violation probability is required to be below $ P_{{\rm{delay}}}^{{\rm{out}}}$, one should limit its incoming data rate to a maximum of $\mu  = {\rm{EC}}(\theta )$.

It is seen from (\ref{frehiutgi}) that $\theta$ is an important parameter, representing the decay rate of the delay violation probability. A smaller $\theta$ corresponds to a slower decay rate, which indicates that the delay requirement is loose, while a larger $\theta$ corresponds to a faster decay rate, which implies that the system is capable of supporting a more stringent delay requirement. In other words, when $\theta\rightarrow0$, an arbitrarily long delay can be tolerated by the system, which corresponds to the capacity studied in Shannon information theory. On the other hand, when $\theta\rightarrow\infty$, this implies that no delay is allowed by the system, which corresponds to the very stringent statistical delay-bound QoS constraint of allowing no delay at all.
\subsection{Problem formulation}
For the user, we assume that the non-coherent joint transmission is adopted as the BBU pool \cite{xuchen2012,xuchen2012wcnc,Chien2016}. Then, the instantaneous service rate of a single frame, denoted by $R(\bm{\nu})$, can be expressed as follows \cite{liangliu2015,chunlonghejsac}:
\vspace{-0.15cm}
\begin{equation}
R(\bm\nu)=T_fB\log_2(1+\sum\nolimits_{i\in\mathcal{I}}p_i(\bm\nu)\alpha_i),\vspace{-0.15cm}
\end{equation}
where $B$ is the system's bandwidth, $\bm\nu  \buildrel \Delta \over = (\bm\alpha ,\theta )$ represents the network condition that includes both the channel's power gains and the EC exponent requirement,  and $p_i(\bm\nu)$ represents the power allocation for RRH $i$ that depends on the network's condition $\bm\nu$.
In this paper, we aim for optimizing the transmit power in order to maximize the EC for the user under two different types of power limitations for each RRH: under an average power constraint and a peak power constraint. The first one is related to the long-term power budget, while the second guarantees that the instantaneous transmit power is below the linear range of  practical power amplifiers. Mathematically, this optimization problem can be formulated as
\begin{subequations}\label{initial-pro}
\begin{align}
\mathop {\max }\limits_{{\left\{ {p_i(\bm\nu)},i\in\mathcal{I} \right\}}}\;\;\;&  - \frac{1}{\theta}{\log}\left(\mathbb{E}_{\bm \alpha} {\left[ {{e^{ - \theta T_fB \log_2\left( {1 + \sum\nolimits_{i \in \mathcal{I}} {{p_i}(\bm\nu){\alpha_i}} } \right)}}} \right]} \right)\\
{\rm{s.t.}}\;\;\;&  \mathbb{E}_{\bm\alpha}\left[{{p_i(\bm\nu)}} \right] \le P^{\text{avg}}_i,\forall i \in \mathcal{I},\\
&0\le{p_i(\bm\nu)} \le {P_i^{\text{peak}}},\forall i \in \mathcal{I},
\end{align}
\end{subequations}
where $\mathbb{E}_{\bm\alpha}\{\cdot \}$ denotes the expectation over $\bm \alpha$, while $P^{\text{avg}}_i$ and $P_i^{{\text{peak}}}$ denote the $i$th RRH's maximum average transmit power constraint and peak transmit power constraint, respectively.

\vspace{-0.2cm}
\section{Optimal Power Allocation Method}\label{genralcase}
By exploiting the fact that $\log (\cdot)$ is a monotonically increasing function, Problem (\ref{initial-pro}) can be equivalently simplified as
\vspace{-0.1cm}
\begin{subequations}
\label{mainproblem}
\begin{align}
\mathop {\min }\limits_{\left\{ {{p_i}({\bm\nu}),\forall i\in\cal I} \right\}}\;&{\mathbb{E}_{\bm{\alpha}}}\left[ {{{\left( {1 + \sum\nolimits_{i \in \mathcal{I}} {{p_i}\left( {\bm\nu} \right){\alpha_i}} } \right)}^{ - \varepsilon (\theta )}}} \right]\\
{\rm{s.t.}}\;\;\;&  \mathbb{E}_{\bm\alpha}\left[{{p_i(\bm\nu)}} \right] \le P^{\text{avg}}_i,\forall i \in \mathcal{I},\label{avergepw}\\
&0\le{p_i(\bm\nu)} \le {P_i^{\text{peak}}},\forall i \in \mathcal{I},\label{peakpow}
\end{align}
\end{subequations}
where we have $\varepsilon \left( \theta  \right) = {{\theta {T_f}B} \mathord{\left/
 {\vphantom {{\theta {T_f}B} {\ln 2}}} \right.
 \kern-\nulldelimiterspace} {\ln 2}}$. In Appendix \ref{convexityproof}, we prove that Problem (\ref{mainproblem}) is a convex optimization problem. Hence,  the Lagrangian duality method can be used to solve Problem (\ref{mainproblem}) with zero optimality gap.

Note that in Problem (\ref{mainproblem}), only the average power constraints are ergodic, while the others are instantaneous power constraints. Similar to \cite{Zhang2009}, we should first introduce the dual variables associated with the average transmit power constraints.  Then, the original problem can be decomposed into several independent subproblems, where each one corresponds to one fading state. In addition, the instantaneous power constraints are enforced for each fading state. Let $\bm\lambda=[\lambda_1, \cdots ,\lambda_I]^T$ represent the nonnegative dual variables associated with the average power constraints.  The Lagrangian function of Problem (\ref{mainproblem}) can be written as
\begin{eqnarray}
\mathcal{L}({\bf{P}}\left({\bm\nu}\right),\bm\lambda)&=&\mathbb{E}_{\bm\alpha}[(1+\sum\nolimits_{i\in\mathcal{I}}p_i(\bm\nu) {\alpha}_i)^{-\varepsilon(\theta)}]\nonumber\\
&&+\sum\nolimits_{i\in\mathcal{I}}\lambda_i(\mathbb{E}_{\bm\alpha}[p_i(\bm\nu)]-{P^{\text{avg}}_i}),\label{lag_a}
\end{eqnarray}
where  ${\bf{P}}\left(\bm\nu  \right) = [ {{p_1}(\bm\nu), \cdots ,{p_I}(\bm\nu)} ]^T$. Let us now define $\mathcal{P}=\{{\bf{P}}\left(\bm\nu \right)|\text{(\ref{peakpow})}\}$. The Lagrange dual function is then given by
\vspace{-0.1cm}
\begin{equation}
g\left( { \bm\lambda } \right) = \mathop {\min }\limits_{{{\bf{P}}\left( \bm\nu \right) \in \mathcal{P}} }\;\mathcal{L}({\bf{P}}\left( \bm\nu \right),\bm\lambda).
\label{dual-function}
\end{equation}
\vspace{-0.1cm}
The dual problem is  defined as
\vspace{-0.1cm}
\begin{equation}
\mathop {\max }\limits_{{\lambda _i} \ge 0,\forall i} g\left( {\bm\lambda} \right).
\label{dual-problem}
\end{equation}

As proven in Appendix \ref{convexityproof}, Problem (\ref{mainproblem}) is a convex optimization problem, which implies that there is no duality gap between the dual problem and the original problem. Thus, solving the dual problem is equivalent to solving the original problem.

To solve the dual problem in (\ref{dual-problem}), we should solve the problem in (\ref{dual-function}) for a fixed $\bm\lambda$, then update the Lagrangian dual variables $\bm\lambda$ by solving the dual problem in (\ref{dual-problem}). Iterate the above two steps until convergence is reached.

\emph{1) Solving the dual function in (\ref{dual-function})}: For a given $\bm\lambda$, we should find the dual function $g(\bm\lambda)$, which can be rewritten as
\vspace{-0.15cm}
\begin{equation}
g(\bm\lambda) = \mathbb{E}\left[ {{\tilde g}\left(\bm\lambda\right)} \right] - \sum\limits_{i \in \mathcal{I}} {{\lambda _i}{P^{\text{avg}}_i}},
\vspace{-0.15cm}
\end{equation}
where we have:
\begin{equation}
\tilde g\left(\bm\lambda\right) = \mathop {\min }\limits_{{{\bf{P}}\left(\bm\nu  \right) \in \mathcal{P}}}\; {\left( {1 + \sum\nolimits_{i \in \mathcal{I}} {{p_i}\left(\bm\nu \right){\alpha_i}} } \right)^{-\varepsilon \left( \theta  \right)}} + \sum\nolimits_{i \in \mathcal{I}} {{\lambda _i}{p_i}\left(\bm\nu  \right)}.
\label{transferedpro}
\end{equation}
Note that $\tilde g\left(\bm\lambda\right)$ can be decoupled into multiple independent subproblems, each corresponding to a specific fading state. Those subproblems have the same structure for each fading state. Hence, to simplify the derivation, $\bm\nu$ is omitted in the following. Each subproblem can be expressed as:
\begin{subequations}\label{transfered-whole}
\begin{align}
\mathop {\min }\limits_{\bf{P}}\;\;\;&{\left( {1 + \sum\nolimits_{i \in \mathcal{I}} {{p_i}{\alpha_i}} } \right)^{ - \varepsilon \left( \theta  \right)}} + \sum\nolimits_{i \in \mathcal{I}} {{\lambda _i}{p_i}} \\
\text{s.t.}\;\;\;&0 \le{p_i} \le {P_i^{\text{peak}}},\forall i \in \mathcal{I}.\label{whole-peak}
\end{align}
\end{subequations}
The above problem is convex. In the following, we obtain the closed-form solution by solving the KKT conditions of Problem (\ref{transfered-whole}).

First, we introduce the nonnegative dual variables of $\mu_i, \forall i$, and $\delta_i, \forall i$ for the associated constraints in (\ref{whole-peak}). The KKT conditions for Problem (\ref{transfered-whole}) can be expressed as
\begin{subequations}
\begin{align}
-\varepsilon(\theta)\!\!\left(\!1\!+\!\sum\limits_{i\in\mathcal{I}}p_i^*\alpha_i\!\right)^{-\varepsilon(\theta)\!-\!1}\alpha_i\!+\!\lambda_i+\mu_i^*-\delta_i^*
\!=0, \forall i\label{kkt-p}\\
\mu_i^*(p_i^*-P_i^{\text{peak}})=0, \forall i\label{kkt-u}\\
\delta_i^*p_i^*=0, \forall i\label{kkt-delta}\\
p_i^*\le P_i^{\text{peak}}, \forall i\label{kkt-pmax}\\
p_i^* \ge 0, \forall i,\label{kkt-active}
\end{align}
\end{subequations}
with $p_i^*\ge 0$, $\delta_i^*\ge 0$, and $\mu_i^*\ge 0$, $\forall i$. Then, we have the following lemma.

\emph{\textbf{Lemma 1}}: For any two arbitrary RRHs $i$ and $j$, if  $p_i^*>0$ and $p_j^*=0$, then the following relationship must hold
\vspace{-0.15cm}
\begin{equation}
\frac{\lambda_i}{\alpha_i}\le\frac{\lambda_j}{\alpha_j}.
\label{pertnuation}
\end{equation}
\vspace{-0.05cm}
\emph{Proof}: Please see Appendix \ref{lemma1}.\hfill\rule{2.7mm}{2.7mm}

Lemma 1 shows that the RRHs associated with a smaller ${{{\lambda _i}} \mathord{\left/
 {\vphantom {{{\lambda _i}} {{\alpha _i}}}} \right.
 \kern-\nulldelimiterspace} {{\alpha _i}}}$ should be assigned a non-zero power, while the RRHs having a larger value should remain silent. Let us hence introduce $\pi$ as a permutation over $\mathcal{I}$, so that we have $\frac{\lambda_{\pi(i)}}{\alpha_{\pi(i)}}\le\frac{\lambda_{\pi(j)}}{\alpha_{\pi(j)}}$, when $i<j, i,j\in \cal I$. Let $\mathcal{I'}\subseteq \mathcal{I}$ be the set of RRHs that transmit at a non-zero power. Then, according to Lemma 1, it can be readily verified that we have $\mathcal{I'}=\{\pi(1),\cdots,\pi(|\mathcal{I'}|)\}$.

\emph{\textbf{Lemma 2}}: There is at most one RRH associated with $0<p_i^*<P_{i}^{\text{peak}}$, and the RRH index is $i=\pi(\mathcal{|I'|})$.

\emph{Proof}: Please see Appendix \ref{lemma2}.\hfill\rule{2.7mm}{2.7mm}

Next, we derive the optimal power allocation of Problem (\ref{transfered-whole}) shown as in Theorem 1.

\emph{\textbf{Theorem 1}}: The optimal solution of Problem (\ref{transfered-whole}) is shown as follows,
\begin{equation}\label{calp}
p_{\!\pi (a)}^* \!\!=\!\! \left\{
  \begin{array}{l}
   P_{\pi(a)}^{\text{peak}},\qquad\qquad\qquad\quad a<|\mathcal{I'}|;\\
   \min\left\{P_{\pi(|\mathcal{I'}|)}^{\text{peak}},{T_{{\pi _{(a)}}}}\right\},\ a=|\mathcal{I'}|;\\
   0,\qquad\qquad\qquad\qquad\quad  a>|\mathcal{I'}|;
   \end{array}
   \right.
\end{equation}
where ${T_{{\pi _{(a)}}}}$ is given by
\begin{equation*}
{T_{{\pi _{(a)}}}} \!=\! \frac{1}{{{\alpha _{\pi (|{\cal I}'|)}}}}\!\left[\! {{{\left(\! {\frac{{{\lambda _{\pi (|{\cal I}'|)}}}}{{\varepsilon (\theta ){\alpha _{\pi (|{\cal I}'|)}}}}} \!\right)}^{\! -\! \frac{1}{{\varepsilon (\theta ) \!+ \!1}}}}\! -\! \sum\limits_{a = 1}^{|{\cal I}'| \!-\! 1}  P_{\pi (a)}^{{\rm{peak}}}{\alpha _{\pi (a)}} \!-\! 1} \!\right]
\end{equation*}
with $|\mathcal{I'}|$ being the largest value of integer $x$, so that we have
\begin{equation}
\frac{\lambda_{\pi(x)}}{\varepsilon(\theta)\alpha_{\pi(x)}}<\left[\sum\limits
_{b=1}^{x-1}P_{\pi(b)}^{\text{peak}}\alpha_{\pi(b)}+1\right]^{-{\varepsilon(\theta)
-1}}.
\label{calmaxI}
\end{equation}

\emph{Proof}: Please refer to Appendix \ref{theorem1}. \hfill\rule{2.7mm}{2.7mm}

It can be seen from (\ref{calp}) that the RRHs with high value of ${{{{\lambda _i}} \mathord{\left/
 {\vphantom {{{\lambda _i}} \alpha }} \right.
 \kern-\nulldelimiterspace} \alpha }_i}$ should be switched off, and the ones with larger values can transmit with peak power. The dual variable  $\lambda _i$ can be regarded as the price, and larger value of $\lambda _i$ will incur higher cost when the RRH is active.

\emph{\textbf{Corollary 1}}: When the number of RRHs is equal to one, i.e. $I=1$, the optimal transmit power is given by
\vspace{-0.15cm}
\begin{equation}\label{hideo}
\begin{array}{l}
p_1^* = \left\{ {\begin{array}{*{20}{l}}
{\ \ 0, \qquad\  {\rm{if}}\;{\alpha _1} < \frac{{{\lambda _1}}}{{\varepsilon (\theta )}};}\\
\begin{array}{l}
F,\;\;\qquad{\rm{if}}\;{\alpha _1} \ge \frac{{{\lambda _1}}}{{\varepsilon (\theta )}}\;{\rm{and}}\;F < P_1^{{\rm{peak}}};
\end{array}\\
\begin{array}{l}
P_1^{{\rm{peak}}},\;\;\; {\rm{if}}\;{\alpha _1} \ge \frac{{{\lambda _1}}}{{\varepsilon (\theta )}}{\rm{and}}\;F\ge P_1^{{\rm{peak}}}.
\end{array}
\end{array}} \right.
\end{array}
\end{equation}
where $F$ is given by
\[F = \frac{1}{{{{\left( {\frac{{{\lambda _1}}}{{\varepsilon (\theta )}}} \right)}^{\frac{1}{{1 + \varepsilon (\theta )}}}}\alpha _1^{\frac{{\varepsilon (\theta )}}{{1 + \varepsilon (\theta )}}}}} - \frac{1}{{{\alpha _1}}}.\]
\emph{Proof}: It is can be readily derived using Theorem 1.\hfill\rule{2.7mm}{2.7mm}

Note that the above result is consistent with the point-to-point result obtained in \cite{cheng2016}.

\emph{2) Solving the dual problem (\ref{dual-problem})}: To solve the dual problem (\ref{dual-problem}), we invoke the subgradient method, which is a simple method of optimizing non-differentiable objective function \cite{Weiyu2006}. The subgradient is required by the subgradient of $g(\cdot)$ at ${\bm{\lambda}}^{(k)}=[\lambda_1^{(k)},\cdots,\lambda_I^{(k)}]^T$ in the $k$th iteration\footnote{According to \cite{Weiyu2006}, a vector ${\bf{d}}$ is a subgradient of $g ({\bm{\lambda}})$ at ${\bm{\lambda}}^{(k)}$, if for all ${\bm{\lambda}}$, $g({\bm{\lambda}} ) \leq g({\bm{\lambda}}^{(k)}) + {{\bf{d}}^T}({\bm{\lambda}}  - {\bm{\lambda}}^{(k)})$ holds.}.

 \emph{\textbf{Theorem 2}}: The subgradient of $g(\cdot)$ at ${\bm{\lambda}}^{(k)}$ in the $k$th iteration is given by
 \begin{equation}\label{subgradient}
   {{\bf{d}}^{(k)}} = {\mathbb{E}_{\bm\alpha} }\left[ {\bf{P}}_{\bm\lambda^{(k)}}^*(\bm\nu) \right] - {{\bf{P}}^{{\rm{avg}}}},
 \end{equation}
 where ${{\bf{P}}_{{\bm \lambda ^{(k)}}}^*}(\bm\nu) = {\left[ {{p_{(1,{\bm \lambda ^{(k)}})}^*(\bm\nu)}, \cdots ,{p_{(I,{\bm\lambda ^{(k)}})}^*(\bm\nu)}} \right]^T}$ is the optimal solution of Problem (\ref{dual-function}) when we have $ \bm\lambda  = {\bm\lambda^{(k)}}$, and ${{\bf{P}}^{{\rm{avg}}}} = {\left[ {P_1^{{\rm{avg}}}, \cdots ,P_I^{{\rm{avg}}}} \right]^T}$.

 \emph{Proof}: Please see Appendix \ref{theorem2}.  \hfill\rule{2.7mm}{2.7mm}

Based on Theorem 2, the Lagrangian dual variables can be updated as
\begin{equation}
{\bm\lambda ^{(k + 1)}} =  {{\bm\lambda ^{(k)}} + {\zeta ^{(k)}}{{\bf{d}}^{(k)}}} ,
\label{dualvariabel}
\end{equation}
where $\zeta^{(k)}$ is the step in the $k^{\rm{th}}$ iteration. The subgradient method is guaranteed to converge if $\zeta^{(k)}$ satisfies $\text{lim}_{k\rightarrow\infty}\zeta^{(k)}=0$ and $\sum\nolimits_{k=1}^\infty \zeta^{(k)}=\infty$\cite{boyd2004convex}. The step size in the simulation section is set as $\zeta^{(k)}=a/k$, where $a$ is the constant step-size parameter.

In summary, the solution of Problem (\ref{initial-pro}) is given in Algorithm 1.

\begin{algorithm}
\caption{Solving Problem $(\rm{\ref{initial-pro}})$}

\textbf{Initialize:}

  \quad Iteration number $k = 0$, ${\bm\lambda ^{(0)}} = [{\lambda _1}^{(0)}, \cdots ,{\lambda _I}^{(0)}]$;


\textbf{Repeat}

  \quad 1. Compute ${\bf{P}}^{(k)}$ with fixed $\bm\lambda^{(k)}$ using (\ref{calp});

  \quad 2. Compute the subgradient ${\bf{d}}^{(k)}$, using (\ref{subgradient});

  \quad 3. Update $\bm\lambda^{(k+1)}$ using (\ref{dualvariabel}), increase $k$ by $1$;

\textbf{Until} convergence

\end{algorithm}

To execute Algorithm 1, there is another issue that has to be tackled, namely how to calculate the average power for each RRH in order to obtain the subgradient ${\bf{d}}^{(k)}$. Given the dual variables ${\bm\lambda^{(k)}}$, the expression of optimal power at each RRH depends on the generations of channel gains $\bm\alpha$. Different generations of $\bm\alpha$ will lead to different orders of ${{{\lambda _i}} \mathord{\left/
 {\vphantom {{{\lambda _i}} {{{\alpha }_i}}}} \right.
 \kern-\nulldelimiterspace} {{{\alpha }_i}}}, i=1,\cdots, I$, and thus different power expressions. Even for a fixed problem-order the power allocation expressions require multiple integrations, which imposes a high computational complexity. As a result, it is a challenge to obtain the expression of average power for each RRH in closed-form for any given ${\bm\lambda^{(k)}}$. In fact, even for the simple case of two RRHs, the average powers generally do not have simple closed-form solutions, as shown in the next section.

To resolve the above issue, we propose an online calculation method for tracking the average power required for each RRH. The main idea is to replace the expectation operator by averaging the power allocations for all samples of channel generations during the fading process. Specifically, let us define $\bar P_i^{(k - 1)}$ as
 \begin{equation}
  \bar P_i^{(k - 1)} = \frac{1}{{k - 1}}\sum\limits_{j = 1}^{k - 1} {p_{(i,\bm\lambda ^{(j)})}^*\left( {{\bm\alpha^{(j)},\theta}} \right)},i=1,\cdots,I,
 \label{onlinecal}
 \end{equation}
where $p_{(i,\bm\lambda ^{(j)})}^*\left( {{\bm\alpha^{(j)},\theta}} \right)$ denotes the optimal power allocation of the $i$th RRH for the $j$th channel generation, and  $\bm\lambda ^{(j)}$ represents the corresponding dual variables. Then, the expectation over ${p_{(i,\bm\lambda ^{(k)})}^*}\left(\bm\nu \right)$ can be approximated as
 \begin{eqnarray}
  &&{\mathbb{E}_{\bm\alpha} }\left[ {p_{(i,\bm\lambda ^{(k)})}^*}\left(\bm\nu \right) \right]\nonumber\\
  &\approx& \bar P_i^{(k)}\nonumber\\
  &=&\frac{{p_{(i,\bm\lambda ^{(k)})}^*\left( {{\bm\alpha^{(k)},\theta}} \right) + (k - 1)\bar P_i^{(k - 1)}}}{k}, \forall i\in \cal I,\label{aprooo}
 \end{eqnarray}
which can be recursively obtained  based on the previous $\bar P_i^{(k - 1)},\forall i$. It is worth noting that this algorithm can be readily applied to the case when the fading statistics are unknown.

Fortunately, for the case of a single RRH\footnote{Note that \cite{cheng2016} did not provide the closed-form expression for the average power for the case of one RRH.}, the average power can be obtained in closed-form, which is given in Appendix \ref{theorem3}. The average power expression for the case of $I=2$ is even more complex, which is extensively studied in the following section.

\vspace{-0.25cm}
\section{Special Case: $I=2$}\label{specialcaseoftwo}
\vspace{-0.1cm}
To show the difficulty of obtaining the closed-form solution for each RRH, we only consider the simplest scenario of two RRHs, i.e., $I=2$. To this end, we first introduce the following lemma, which will be used in the ensuing derivations.

\begin{figure}
\centering
\includegraphics[width=3.7in]{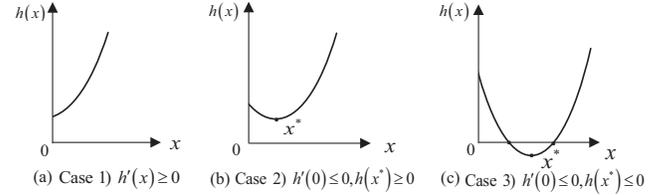}
\vspace{-0.6cm}
\caption{The properties of the function $h(x)=(1+ax)^b-cx$.}
\label{caseI1}
\vspace{-0.5cm}
\end{figure}

\textbf{\emph{Lemma 3:}} Let us define the function $h(x)=(1+ax)^b-cx$, where $a>0,c>0,b>1$. Then, in the region of $x\geq 0$, there are only three possible curves for the function $h(x)$, which are shown in Fig.~\ref{caseI1}. The conditions for each case are given as follows:
\begin{enumerate}
  \item Case 1: $ab-c\geq0$;
  \item Case 2: $ab-c<0$ and ${\left( {\frac{c}{{ab}}} \right)^{{1 \mathord{\left/
 {\vphantom {1 {(b - 1)}}} \right.
 \kern-\nulldelimiterspace} {(b - 1)}}}}(1 - b) + b \ge 0$;
  \item Case 3: $ab-c<0$ and ${\left( {\frac{c}{{ab}}} \right)^{{1 \mathord{\left/
 {\vphantom {1 {(b - 1)}}} \right.
 \kern-\nulldelimiterspace} {(b - 1)}}}}(1 - b) + b < 0$;
\end{enumerate}
 \emph{Proof}: Please see Appendix \ref{lemma4}.\hfill\rule{2.7mm}{2.7mm}

Based on Lemma 3, we commence by deriving the expression of the average transmit power. According to Theorem 1, there are three different possible cases for $I=2$: 1) $\left| {{\cal I}'} \right|=0$; 2) $\left| {{\cal I}'} \right|=1$; 3) $\left| {{\cal I}'} \right|=2$. Obviously, the average power contributed by the first case is zero, hence we only consider the latter two cases, which are discussed in the following two subsections. Without loss of generality (w.l.o.g), we assume that
\begin{equation}\label{paixu}
\frac{\lambda_1}{{\alpha}_1}\le\frac{\lambda_2}{{\alpha}_2}
\end{equation}
holds. The case for $\frac{\lambda_1}{{\alpha}_1}>\frac{\lambda_2}{{\alpha}_2}$ can be similarly derived.
\vspace{-0.3cm}
\subsection{Scenario 1: There is only one RRH that transmits at a non-zero power, i.e., $\left| {{\cal I}'} \right|=1$}\label{dee}
Since we have assumed that the inequality (\ref{paixu}) holds, only  RRH 1 is transmitting at a non-zero power and RRH 2 remains silent in this case. According to Theorem 1, the conditions for $\left| {{\cal I}'} \right|=1$ are given by
\begin{align}
\frac{{{\lambda _2}}}{{\varepsilon \left( \theta  \right){{ \alpha }_2}}} \ge {\left( {P_1^{\rm{peak} }{{ \alpha }_1} + 1} \right)^{ - \varepsilon \left( \theta  \right) - 1}},\label{one}\\
\frac{{{\lambda _1}}}{{\varepsilon \left( \theta  \right){{ \alpha }_1}}} < 1,\label{two}
\end{align}
and only RRH 1 transmits at non-zero power, which is given by
\begin{equation}\label{RRH1power}
 p_1^* = \min \left\{ {P_1^{\text{peak} },\frac{1}{{{{ \alpha }_1}}}\left[ {{{\left( {\frac{{{\lambda _1}}}{{\varepsilon \left( \theta  \right){{ \alpha }_1}}}} \right)}^{ - \frac{1}{{\varepsilon \left( \theta  \right) + 1}}}} - 1} \right]} \right\}.
\end{equation}
It may be readily seen that, there are two possible values of $p_1^*$, and the conditions for each value will be discussed as follows.

\subsubsection{$p_1^*=P_1^{\text{peak} }$}

In this case, the following condition should be satisfied:
\begin{equation}\label{pmaxcondi}
\frac{1}{{{{ \alpha }_1}}}\left[ {{{\left( {\frac{{{\lambda _1}}}{{\varepsilon \left( \theta  \right){{ \alpha }_1}}}} \right)}^{ - \frac{1}{{\varepsilon \left( \theta  \right) + 1}}}} - 1} \right] \ge P_1^{\text{peak} },
\end{equation}
which is equivalent to
\begin{equation}\label{equooo}
 {\left( {1 + P_1^{\text{peak} }{{ \alpha }_1}} \right)^{1 + \varepsilon \left( \theta  \right)}} - \frac{{\varepsilon \left( \theta  \right)}}{{{\lambda _1}}}{{ \alpha }_1} \le 0.
\end{equation}
Notice that the left hand side of (\ref{equooo}) is in the form of $h(x)$ defined in Lemma 3, with $x = {{ \alpha }_1}$, $a,b,c$ given by
\begin{equation}\label{abc}
  a = P_1^{\text{peak} },b = 1 + \varepsilon \left( \theta  \right),c = {{\varepsilon \left( \theta  \right)} \mathord{\left/
 {\vphantom {{\varepsilon \left( \theta  \right)} {{\lambda _1}}}} \right.
 \kern-\nulldelimiterspace} {{\lambda _1}}}.
\end{equation}

To guarantee that there exists a positive ${ \alpha }_1$, the function $h(x)$ should be reminiscent of Fig.~\ref{caseI1}-(c), and the conditions for Case 3 in Lemma 3 should be satisfied. Otherwise, $p_1^*$ is not equal to $P_1^{\text{peak} }$. In the following, we assume that the conditions are satisfied. Let us denote the solutions of $h({{ \alpha }_1})=0$ as $ \alpha _1^l$ and $ \alpha _1^u$, where $ \alpha _1^l< \alpha _1^u$. These solutions can be readily obtained by using the classic bisection method. Then, condition (\ref{equooo}) is equivalent to
\begin{equation}\label{conditongamma}
 \alpha _1^l \le {{ \alpha }_1} \le  \alpha _1^u.
\end{equation}
By combining (\ref{paixu}) and (\ref{one}), we obtain the feasible region of ${\alpha}_2$ in the form of:
\begin{eqnarray}
 0 \le {{ \alpha }_2} &\le& \min \left( {\frac{{{\lambda _2}}}{{{\lambda _1}}}{{ \alpha }_1},\frac{{{\lambda _2}}}{{\varepsilon \left( \theta  \right)}}{{\left( {1 + P_1^{\text{peak} }{{ \alpha }_1}} \right)}^{1 + \varepsilon \left( \theta  \right)}}} \right)\nonumber\\
 &=& \frac{{{\lambda _2}}}{{\varepsilon \left( \theta  \right)}}{\left( {1 + P_1^{\text{peak} }{{ \alpha }_1}} \right)^{1 + \varepsilon \left( \theta  \right)}},\label{gammme}
\end{eqnarray}
where the last equality holds by using (\ref{equooo}). Similarly, by combining (\ref{two}) and (\ref{conditongamma}), we obtain the feasible region of ${\alpha}_1$ as:
\begin{equation}\label{callll}
 \max \left( { \alpha _1^l,\frac{{{\lambda _1}}}{{\varepsilon \left( \theta  \right)}}} \right) \le {{ \alpha }_1} \le  \alpha _1^u.
\end{equation}
In this context, we prove that ${{{\lambda _1}} \mathord{\left/
 {\vphantom {{{\lambda _1}} {\varepsilon \left( \theta  \right)}}} \right.
 \kern-\nulldelimiterspace} {\varepsilon \left( \theta  \right)}} < \alpha _1^l$ in Appendix \ref{daxiaorelation}. Hence, the feasible region of ${\alpha}_1$ is given by $\alpha _1^l \le {\alpha _1} \le \alpha _1^u$.

Based on the above discussions, we obtain the conditions for $p_1^*=P_1^{\text{peak} }, p_2^*=0$ as follows:
\begin{equation}\label{aii}
  {\rm{C1:}} ab-c<0, {\left( {\frac{c}{{ab}}} \right)^{{1 \mathord{\left/
 {\vphantom {1 {(b - 1)}}} \right.
 \kern-\nulldelimiterspace} {(b - 1)}}}}(1 - b) + b < 0,
\end{equation}
where $a,b,c$ are given in (\ref{abc}).

 If Condition C1 in (\ref{aii}) is satisfied, the average power assigned to RRH 1 in this case is given by
 \begin{equation}\label{T11RRH1}
 T_{{\rm{RRH}}_1}^{{\rm{C1}}} \!\!\buildrel \Delta \over =\!\! \int_{ \alpha _1^l}^{\alpha _1^u}\!\!\!\! {\int_0^{\frac{{{\lambda _2}}}{{\varepsilon \left( \theta  \right)}}{{\left( {1 + P_1^{\text{peak} }{{ \alpha }_1}} \right)}^{1 + \varepsilon \left( \theta  \right)}}}\!\!\!\!\!\!\!\! {P_1^{\text{peak} } f\left( {{{ \alpha }_1}} \right) f\left( {{{ \alpha }_2}} \right)d{{ \alpha }_2}d{{ \alpha }_1}} }.
 \end{equation}
 By substituting the PDF of ${ \alpha }_2$ in (\ref{pdfex}) into (\ref{T11RRH1}), (\ref{T11RRH1}) can be simplified to
\begin{equation}\label{simpli}
\begin{array}{l}
T_{{\rm{RR}}{{\rm{H}}_1}}^{{\rm{C1}}} = \\
\frac{{P_1^{{\rm{peak}}}}}{{\Gamma (m)}}\int_{\alpha _1^l}^{\alpha _1^u} {f({\alpha _1})\gamma \left( {m,\frac{{m{\lambda _2}}}{{{{\bar \alpha }_2}\varepsilon \left(\!\! \theta  \right)}}{{\left( {1 + P_1^{{\rm{peak}}}{\alpha _1}} \right)}^{1 + \varepsilon \left( \theta  \right)}}} \!\!\right)d{\alpha _1}} .
\end{array}
\end{equation}
Unfortunately, the closed-form expression of $T_{{\rm{RRH}}_1}^{{\rm{C1}}}$ cannot be obtained even for the special case of $m=1$. However, the value of  $T_{{\rm{RRH}}_1}^{{\rm{C1}}}$ can be obtained at a good accuracy by using the numerical integration function of Matlab.

\subsubsection{$p_1^* = \frac{1}{{{{ \alpha }_1}}}\left[ {{{\left( {\frac{{{\lambda _1}}}{{\varepsilon \left( \theta  \right){{ \alpha }_1}}}} \right)}^{ - \frac{1}{{\varepsilon \left( \theta  \right) + 1}}}} - 1} \right]$}

In this case, the following condition should be satisfied:
\begin{equation}\label{pmidcone}
\frac{1}{{{{ \alpha }_1}}}\left[ {{{\left( {\frac{{{\lambda _1}}}{{\varepsilon \left( \theta  \right){{ \alpha }_1}}}} \right)}^{ - \frac{1}{{\varepsilon \left( \theta  \right) + 1}}}} - 1} \right] < P_1^{\text{peak} },
\end{equation}
which leads to $h({{ \alpha }_1})>0$ with $a,b,c$ defined in (\ref{abc}). As seen from Fig.~\ref{caseI1}, when the first two conditions in Lemma 3 are satisfied, the inequality (\ref{pmidcone}) holds for any ${ \alpha }_1\geq0$. When the third condition in Lemma 3 is satisfied, the inequality (\ref{pmidcone}) holds when $0 \le { \alpha _1} \le \alpha _1^l$ and ${ \alpha _1} \ge  \alpha _1^u$, where $\alpha _1^l$ and $\alpha _1^u$ ($\alpha _1^l<\alpha _1^u$ ) are the solutions of $h(\alpha _1)=0$ with $a,b,c$ defined in (\ref{abc}).

Now, we first assume that the first two conditions in Lemma 3 are satisfied
 \begin{equation}\label{fff}
{\rm{C2:}}ab - c \ge 0, {\rm{or}}\  ab-c<0\  {\rm{and}}\  {\left( {\frac{c}{{ab}}} \right)^{{1 \mathord{\left/
 {\vphantom {1 {(b - 1)}}} \right.
 \kern-\nulldelimiterspace} {(b - 1)}}}}(1 - b) + b \geq0,
\end{equation}
where $a,b,c$ are given in (\ref{abc}). Then, the condition in (\ref{pmidcone}) can be neglected. According to (\ref{paixu}) and (\ref{one}), the feasible region of ${ \alpha }_2$ is given by
\begin{equation}\label{afaftow}
 0 \le {{ \alpha }_2} \le \min \left( {\frac{{{\lambda _2}}}{{{\lambda _1}}}{{ \alpha }_1},\frac{{{\lambda _2}}}{{\varepsilon \left( \theta  \right)}}{{\left( {1 + P_1^{\text{peak} }{{ \alpha }_1}} \right)}^{1 + \varepsilon \left( \theta  \right)}}} \right) = \frac{{{\lambda _2}}}{{{\lambda _1}}}{ \alpha _1},
\end{equation}
where $(\ref{pmidcone})$ is used in the last equality. From (\ref{two}), the feasible region of ${ \alpha }_1$ is given by ${ \alpha _1} \ge \frac{{{\lambda _1}}}{{\varepsilon \left( \theta  \right)}}$. As a result, the average power of RRH 1 contributed by the case when condition C2 is satisfied is given by
\begin{align}
T_{{\rm{RRH}}_1}^{{\rm{C2}}} &= \int_{\frac{\lambda _1}{{\varepsilon \left( \theta  \right)}}}^\infty  {\int_0^{\frac{\lambda _2}{\lambda _1}{\alpha }_1} {p_1^*f\left( {{\alpha _1}} \right)f\left( {{\alpha _2}} \right)d{\alpha _2}d{\alpha _1}} },\nonumber\\
&= \frac{1}{{\Gamma \left( m \right)}}\int_{\frac{{{\lambda _1}}}{{\varepsilon \left( \theta  \right)}}}^\infty  {p_1^*\gamma \left( {m,\frac{{m{\lambda _2}{\alpha _1}}}{{{{\bar \alpha }_2}{\lambda _1}}}} \right)f\left( {{\alpha _1}} \right)d{\alpha _1}}  .\label{conditw}
\end{align}

Fortunately, when $m$ is an integer, the closed-form expression of $T_{{\rm{RRH}}_1}^{{\rm{C2}}}$ can be obtained. Let us define
\begin{equation}\label{ff}
  \begin{array}{l}
U \buildrel \Delta \over = \frac{{{\lambda _1}}}{{\varepsilon \left( \theta  \right)}},V \buildrel \Delta \over = \frac{1}{{1 + \varepsilon \left( \theta  \right)}},W \buildrel \Delta \over = \frac{{m{\lambda _2}}}{{{{\bar \alpha }_2}{\lambda _1}}},\\
Z \buildrel \Delta \over = \frac{{{{\left( {m/{{\bar \alpha }_1}} \right)}^m}}}{{(m - 1)!}},Y \buildrel \Delta \over = \frac{{m{\lambda _2}}}{{{{\bar \alpha }_2}{\lambda _1}}} + \frac{m}{{{{\bar \alpha }_1}}}.
\end{array}
\end{equation}
If $m=1$, the Nakagami-$m$ channel reduces to the Rayleigh channel, and the average power of RRH 1 in (\ref{conditw}) can be simplified to:
\begin{equation}\label{SIMPLECASE}
\begin{array}{l}
  T_{{\rm{RR}}{{\rm{H}}_1}}^{{\rm{C2}}} = \frac{{\bar \alpha _1^{V-1}\Gamma \left( {V,{U \mathord{\left/
 {\vphantom {U {{{\bar \alpha }_1}}}} \right.
 \kern-\nulldelimiterspace} {{{\bar \alpha }_1}}}} \right)}}{{{U^V}}} + \frac{1}{{{{\bar \alpha }_1}}}{\rm{Ei}}\left( { - \frac{U}{{{{\bar \alpha }_1}}}} \right) \\
\qquad\qquad - \frac{{\Gamma \left( {V,UY} \right)}}{{{{\bar \alpha }_1}{Y^V}{U^V}}} - \frac{1}{{{{\bar \alpha }_1}}}{\rm{Ei}}\left( { - UY} \right).
 \end{array}
\end{equation}
 If $m$ is an integer that is larger than one, i.e., $m\geq2$,  we can obtain the closed-form expression of $T_{{\rm{RRH}}_1}^{{\rm{C2}}}$:
\begin{equation}\label{FhaDE}
  T_{{\rm{RRH}}_1}^{{\rm{C2}}} = {J_1} + {J_2} - {J_3} - {J_4},
\end{equation}
where $J_1$, ${J_2}$, ${J_3}$  and ${J_4}$ are respectively given by
\begin{equation}\label{J1J2}
  \begin{array}{l}
{J_1} = \frac{{\bar \alpha _1^{V - 1}\Gamma \left( {V + m - 1,\frac{{mU}}{{{{\bar \alpha }_1}}}} \right)}}{{(m - 1)!{m^{V - 1}}{U^V}}},\\
{J_2} = Z\sum\limits_{l = 0}^{m - 1} {\frac{{{W^l}}}{{l!{Y^{l + m - 1}}}}\Gamma \left( {l + m - 1,YU} \right)} ,
\end{array}
\end{equation}
\begin{equation}\label{J3J4}
 \begin{array}{l}
{J_3} = \frac{Z}{{{U^V}}}\sum\limits_{l = 0}^{m - 1} {\frac{{{W^l}}}{{l!{Y^{l + V + m - 1}}}}\Gamma \left( {l + V + m - 1,YU} \right)} ,\\
{J_4} = \frac{m}{{{{\bar \alpha }_1}(m - 1)!}}\Gamma \left( {m - 1,\frac{{mU}}{{{{\bar \alpha }_1}}}} \right).
\end{array}
\end{equation}
The details of the derivations can be found in Appendix \ref{derivaitons}.

Next, we consider the case, where Condition 3) in Lemma 3 is satisfied. According to the condition in (\ref{pmidcone}), the feasible region of ${ \alpha }_1$ is $0 \le { \alpha _1} \le  \alpha _1^l$ and ${ \alpha _1} \ge  \alpha _1^u$. Additionally, from (\ref{two}), ${{ \alpha }_1} \geq U$ should hold. According to Appendix \ref{daxiaorelation},  $U\leq  \alpha _1^l$ always holds. Hence, the overall feasible region is $ \alpha _1^l \ge {{ \alpha }_1} \geq U$ and ${ \alpha _1} \ge  \alpha _1^u$.  Furthermore, the feasible region of ${ \alpha }_2$ is the same as in (\ref{afaftow}). As a result, the expression of the average power for RRH 1 contributed in this case can be similarly obtained as in (\ref{conditw}), except for the different integration intervals for $\alpha_1$. The expression for the special case, when $m$ is an integer can  be similarly obtained.

Let us now define the following function
\[F(x) = \frac{1}{{\Gamma \left( m \right)}}\int_x^\infty  {\frac{1}{{{\alpha _1}}}\left[ {{{\left( {\frac{{{\alpha _1}}}{U}} \right)}^V} - 1} \right]\gamma \left( {m,W{\alpha _1}} \right){f({\alpha _1})}d{\alpha _1}}. \]
Then, $T_{{\rm{RRH}}_1}^{{\rm{C2}}}$ given in (\ref{conditw}) is equal to $T_{{\rm{RRH}}_1}^{{\rm{C2}}}=F(U)$.

If the following condition is satisfied:
\begin{equation}\label{afza}
{\rm{C3}}:  ab-c<0 \ {\rm{and}}\  {\left( {\frac{c}{{ab}}} \right)^{{1 \mathord{\left/
 {\vphantom {1 {(b - 1)}}} \right.
 \kern-\nulldelimiterspace} {(b - 1)}}}}(1 - b) + b < 0,
\end{equation}
with $a,b,c$  given in (\ref{abc}), the average power for RRH 1 contributed under this condition is
\begin{equation}\label{tejdeo}
 T_{{\rm{RRH}}_1}^{{\rm{C3}}} = F\left( U \right) - F\left( { \alpha _1^l} \right) + F\left( { \alpha _1^u} \right).
\end{equation}
Note that Condition C3 is the same as Condition C1, but the power allocation for RRH 1 is different.

\vspace{-0.4cm}
\subsection{Scenario 2: Both RRHs are transmitting at a non-zero power, i.e., $\left| {{\cal I}'} \right|=2$}
In this case, RRH 1 will transmit with full power, i.e., $p_1^*=P_1^{\rm{peak}}$, and RRH 2 will transmit at a non-zero power. According to Theorem 1, the following condition should be satisfied:
\begin{equation}\label{doiewhdo}
 \frac{{{\lambda _2}}}{{\varepsilon \left( \theta  \right){{ \alpha }_2}}} < {\left( {1 + P_1^{\text{peak} }{{ \alpha }_1}} \right)^{ - \varepsilon \left( \theta  \right) - 1}}
\end{equation}
and the transmit power of RRH 2 is given by
\begin{equation}\label{hoiew}
  p_2^* = \min \left\{\!\! {P_2^{\text{peak} },\frac{1}{{{{ \alpha }_2}}}\!\!\left[\!\! {{{\left( {\frac{{{\lambda _2}}}{{\varepsilon \left( \theta  \right){{ \alpha }_2}}}} \right)}^{ - \frac{1}{{1 \!+ \!\varepsilon \left( \theta  \right)}}}} \!- P_1^{\text{peak} }{{ \alpha }_1} - 1}\!\! \right]}\!\! \right\}.
\end{equation}
The conditions for each value of $p_2^*$ will be discussed as follows.

\subsubsection{$p_2^*=P_2^{\text{peak} }$}
In this case, the following condition should be satisfied:
\begin{equation}\label{dwdaew}
  P_2^{\text{peak} } \le \frac{1}{{{{ \alpha }_2}}}\left[ {{{\left( {\frac{{{\lambda _2}}}{{\varepsilon \left( \theta  \right){{ \alpha }_2}}}} \right)}^{ - \frac{1}{{1 + \varepsilon \left( \theta  \right)}}}} - P_1^{\text{peak} }{{ \alpha }_1} - 1} \right].
\end{equation}
Combining conditions (\ref{paixu}), (\ref{doiewhdo}) and (\ref{dwdaew}), we can obtain the feasible region of ${ \alpha }_1$ as follows
\begin{equation}\label{joifr}
  \frac{{{\lambda _1}}}{{{\lambda _2}}}{{ \alpha }_2} \!\le\! {{ \alpha }_1} \!< \!\frac{1}{{P_1^{\text{peak} }}}\left[ {{{\left( {\frac{{\varepsilon \left( \theta  \right){{ \alpha }_2}}}{{{\lambda _2}}}} \right)}^{\frac{1}{{1 + \varepsilon \left( \theta  \right)}}}} \!-\! P_2^{\text{peak} }{{ \alpha }_2} -\! 1} \right]\triangleq A.
\end{equation}
To guarantee that we have a nonempty set of ${ \alpha }_1$, the following condition should be satisfied:
\begin{equation}\label{fhyt}
 {\left( {1 + P_2^{\text{peak} }{{ \alpha }_2} + \frac{{{\lambda _1}}}{{{\lambda _2}}}P_1^{\text{peak} }{{ \alpha }_2}} \right)^{1 + \varepsilon \left( \theta  \right)}} < \frac{{\varepsilon \left( \theta  \right)}}{{{\lambda _2}}}{{ \alpha }_2},
\end{equation}
which is in the form of function $h(x)$ defined in Lemma 3 with $x={ \alpha }_2$, and $a,b,c$ given by
\begin{equation}\label{dsef}
a = P_2^{\text{peak} } + \frac{{{\lambda _1}}}{{{\lambda _2}}}P_1^{\text{peak} },b = 1 + \varepsilon \left( \theta  \right),c = \frac{{\varepsilon \left( \theta  \right)}}{{{\lambda _2}}}.
\end{equation}
To guarantee the existence of a positive ${ \alpha }_2$, the graphical curve of the function $h({ \alpha }_2)$ should be similar to that in Fig.~\ref{caseI1}-c, and the conditions are given by
\begin{equation}\label{dfdser}
 {\rm{C4:}} ab-c<0, {\rm{and}}\  {\left( {\frac{c}{{ab}}} \right)^{{1 \mathord{\left/
 {\vphantom {1 {(b - 1)}}} \right.
 \kern-\nulldelimiterspace} {(b - 1)}}}}(1 - b) + b < 0,
\end{equation}
where $a,b,c$ are given in (\ref{dsef}). Let us denote the solutions of $h({ \alpha }_2)=0$ as $ \alpha _2^l$ and $ \alpha _2^u$, where we have $ \alpha _2^l< \alpha _2^u$. Then, the feasible region of $ \alpha _2$ is given by $ \alpha _2^l< \alpha _2< \alpha _2^u$.

Under Condition C4, the average powers of RRH 1 and RRH 2 are respectively calculated as
\begin{align}
&T_{{\rm{RR}}{{\rm{H}}_{\rm{1}}}}^{{\rm{C4}}} =\nonumber \\
&\frac{{P_{\rm{1}}^{{\rm{peak}}}}}{{\Gamma \left( m \right)}}\int_{\alpha _2^l}^{\alpha _2^u} {f({\alpha _2})\left[ {\gamma \left( {m,\frac{{mA}}{{{{\bar \alpha }_1}}}} \right) - \gamma \left( {m,\frac{{m{\lambda _1}{\alpha _2}}}{{{{\bar \alpha }_1}{\lambda _2}}}} \right)} \right]} d{\alpha _2},\label{r1}\\
&T_{{\rm{RR}}{{\rm{H}}_2}}^{{\rm{C4}}} = \frac{{T_{{\rm{RR}}{{\rm{H}}_{\rm{1}}}}^{{\rm{C4}}}P_2^{\rm{peak} }}}{{P_1^{\rm{peak} }}},\label{r2}
\end{align}
where $A$ is defined in (\ref{joifr}).

\subsubsection{$p_2^*=\frac{1}{{{{ \alpha }_2}}}\left[ {{{\left( {\frac{{{\lambda _2}}}{{\varepsilon \left( \theta  \right){{ \alpha }_2}}}} \right)}^{ - \frac{1}{{1 + \varepsilon \left( \theta  \right)}}}} - P_1^{\text{peak} }{{ \alpha }_1} - 1} \right]$}

In this case, the following condition should be satisfied:
\begin{equation}\label{dew}
  P_2^{\text{peak} } > \frac{1}{{{{ \alpha }_2}}}\left[ {{{\left( {\frac{{{\lambda _2}}}{{\varepsilon \left( \theta  \right){{ \alpha }_2}}}} \right)}^{ - \frac{1}{{1 + \varepsilon \left( \theta  \right)}}}} - P_1^{\text{peak} }{{ \alpha }_1} - 1} \right].
\end{equation}
By combining (\ref{paixu}), (\ref{doiewhdo}) and (\ref{dew}), the feasible region of ${ \alpha }_1$ can be obtained as follows:
\begin{equation}\label{frsf}
B\!\triangleq\!\frac{1}{{P_1^{\text{peak} }}}\left[\! {{{\left( {\frac{{\varepsilon \left( \theta  \right){{ \alpha }_2}}}{{{\lambda _2}}}} \right)}^{\frac{1}{{1 \!+\! \varepsilon \left( \theta  \right)}}}}\! -\! 1} \!\right] \!>\! {{ \alpha }_1} \!\ge\! \max \left\{ {\frac{{{\lambda _1}}}{{{\lambda _2}}}{{ \alpha }_2},A} \right\},
\end{equation}
where $A$ is defined in (\ref{joifr}). Note that $B>A$ always holds. Hence, to guarantee that there exists a feasible ${ \alpha }_1$, the following condition should be satisfied:
\begin{equation}\label{dewaa}
 0 \ge {\left( {1 + \frac{{{\lambda _1}}}{{{\lambda _2}}}P_1^{\text{peak} }{{ \alpha }_2}} \right)^{1 + \varepsilon \left( \theta  \right)}} - \frac{{\varepsilon \left( \theta  \right){{ \alpha }_2}}}{{{\lambda _2}}}.
\end{equation}
Again, the right hand side of (\ref{dewaa}) is in the form of the function $h(x)$ defined in Lemma 3, with $x={{{ \alpha }_2}}$, and $a,b,c$ are given by
\begin{equation}\label{devarfv}
  a = \frac{{{\lambda _1}}}{{{\lambda _2}}}P_1^{\text{peak} },b = 1 + \varepsilon \left( \theta  \right),c = \frac{{\varepsilon \left( \theta  \right)}}{{{\lambda _2}}}.
\end{equation}
To guarantee that there exists a positive ${ \alpha }_2$, the third condition in Lemma 3 should be satisfied:
\begin{equation}\label{oignjn}
 {\rm{C_X:}} ab-c<0\  {\rm{and}}\  {\left( {\frac{c}{{ab}}} \right)^{{1 \mathord{\left/
 {\vphantom {1 {(b - 1)}}} \right.
 \kern-\nulldelimiterspace} {(b - 1)}}}}(1 - b) + b < 0,
\end{equation}
where $a,b,c$ are given in (\ref{devarfv}). When $a,b,c$ are given in (\ref{devarfv}), we can denote the solutions of ${h\left( {{{ \alpha }_2}} \right)}=0$ as
${ \tilde \alpha _2^l}$ and ${ \tilde \alpha _2^u}$ with ${ \tilde \alpha _2^l}<\tilde \alpha _2^u$. Hence, the feasible region of $\alpha_2$ is given by ${ \tilde\alpha _2^l}<\alpha_2<\tilde\alpha _2^u$.

The remaining task is to determine the lower bound of ${ \alpha }_1$ as shown in (\ref{frsf}).

\emph{Case I: }If the following condition is satisfied:
\begin{equation}\label{deef}
 \frac{{{\lambda _1}}}{{{\lambda _2}}}{\alpha _2} \ge A,
\end{equation}
the feasible region of  ${ \alpha }_1$ is given by
\begin{equation}\label{dewa}
  B > {\alpha _1} \ge \frac{{{\lambda _1}}}{{{\lambda _2}}}{\alpha _2},
\end{equation}
where $B$ is defined in (\ref{frsf}). The inequality (\ref{deef}) can be rewritten as
\begin{equation}\label{dewdA}
  {\left( {1 + P_2^{{\rm{peak}}}{\alpha _2} + \frac{{{\lambda _1}P_1^{{\rm{peak}}}}}{{{\lambda _2}}}{\alpha _2}} \right)^{1 + \varepsilon \left( \theta  \right)}} \ge \frac{{\varepsilon \left( \theta  \right){\alpha _2}}}{{{\lambda _2}}},
\end{equation}
which is equivalent to $h(\alpha _2)\geq 0$ with $a$, $b$ and $c$ given in (\ref{dsef}).

As seen from Fig.~\ref{caseI1}, when the first two conditions in Lemma 3 are satisfied, the inequality (\ref{dewdA}) holds for any $\alpha _2>0$. Combining this with the condition (\ref{dewaa}), the feasible region of $\alpha _2$ is given by ${ \tilde\alpha _2^l}<\alpha _2<\tilde\alpha _2^u$, where $\tilde\alpha _2^l$ and $\tilde\alpha _2^u$ are the solutions of $h(\alpha _2)=0$ with $a,b,c$ defined in (\ref{devarfv}). Define the following condition with $a,b,c$ defined in (\ref{dsef})\footnote{Under Condition C5, $a,b,c$ are defined in (\ref{devarfv}).}:
\begin{equation}\label{hoidehi}
 \begin{array}{l}
{\rm{C}}5:{{\rm{C}}_{\rm{X}}}\;{\rm{and}}\;ab - c \ge 0,{\rm{or}}\;{{\rm{C}}_{\rm{X}}},\;ab - c < 0\\
\;\qquad{\rm{and}}\;{\left( {\frac{c}{{ab}}} \right)^{1/(b - 1)}}\left( {1 - b} \right){\rm{ + }}b \ge 0,
\end{array}
\end{equation}
and $C$ as
\begin{equation}\label{dewnioij}
 C \triangleq\frac{1}{{{\alpha _2}}}\left[ {{{\left( {\frac{{\varepsilon \left( \theta  \right){\alpha _2}}}{{{\lambda _2}}}} \right)}^{\frac{1}{{1 + \varepsilon \left( \theta  \right)}}}} - 1} \right].
\end{equation}
Under Condition C5, the average power of RRH 1 and RRH 2 are respectively given by
\begin{align}
&T_{{\rm{RR}}{{\rm{H}}_1}}^{{\rm{C5}}} = \nonumber\\
&\frac{{P_{\rm{1}}^{{\rm{peak}}}}}{{\Gamma \left( m \right)}}\int_{\tilde\alpha _2^l}^{\tilde\alpha _2^u} {\left[ {\gamma \left( {m,\frac{{mB}}{{{{\bar \alpha }_1}}}} \right) - \gamma \left( {m,\frac{{m{\lambda _1}}}{{{{\bar \alpha }_1}{\lambda _2}}}{\alpha _2}} \right)} \right]f({\alpha _2})} d{\alpha _2},\\
&T_{{\rm{RR}}{{\rm{H}}_2}}^{{\rm{C5}}} = {{ J}_1} - {{ J}_2}
\end{align}
with
\[\begin{array}{l}
{J_1} = \frac{1}{{\Gamma \left( m \right)}}\int_{\tilde\alpha _2^l}^{\tilde\alpha _2^u} C \!\left(\!\! {\gamma \left( {m,\frac{{mB}}{{{{\bar \alpha }_1}}}} \right) \!- \!\gamma \left( {m,\frac{{m{\lambda _1}}}{{{{\bar \alpha }_1}{\lambda _2}}}{\alpha _2}} \right)}\!\! \right)f\left( {{\alpha _2}} \right)d{\alpha _2}\\
{J_2} = \frac{{{{\bar \alpha }_1}P_1^{{\rm{peak}}}}}{{m\Gamma \left( m \right)}}\int_{\tilde \alpha _2^l}^{\tilde \alpha _2^u} {\frac{1}{{{\alpha _2}}}\left( {E - G} \right)f\left( {{\alpha _2}} \right)} d{\alpha _2},
\end{array}\]
where $E$ and $G$ are respectively given by
\[E = \gamma \left( {m + 1,\frac{{mB}}{{{{\bar \alpha }_1}}}} \right),G = \gamma \left( {m + 1,\frac{{m{\lambda _1}}}{{{{\bar \alpha }_1}{\lambda _2}}}{\alpha _2}} \right)\]
with $B$ defined in (\ref{frsf}), ${\tilde\alpha _2^l}$ and ${\tilde\alpha _2^u}$ are the solutions of $h(\alpha _2)=0$ with $a,b,c$ defined in (\ref{devarfv}).

On the other hand, when the third condition in Lemma 3 is satisfied, inequality (\ref{dewdA}) holds for $0 \le {\alpha _2} \le \alpha _2^l$ or ${\alpha _2} \ge  \alpha _2^u$, where $ \alpha _2^l$ and $ \alpha _2^u$ are the solutions of $h(\alpha _2)=0$ with $a,b,c$ defined in (\ref{dsef}). Additionally, it is easy to verify that the curve of $h(\alpha _2)$ associated with $a,b,c$ defined in (\ref{dsef}) is above the curve of $h(\alpha _2)$ with $a,b,c$ defined in (\ref{devarfv}). Hence, the following relations hold:
\vspace{-0.2cm}
\begin{equation}\label{relation}
 \tilde  \alpha _2^l < \alpha _2^l <  \alpha _2^u < \tilde \alpha _2^u.
 \vspace{-0.2cm}
\end{equation}
Combining (\ref{relation}) with condition (\ref{dewaa}), the feasible region of $\alpha _2$ is given by
\vspace{-0.2cm}
\begin{equation}\label{oihe}
 \tilde \alpha _2^l < {\alpha _2} <  \alpha _2^l,  \alpha _2^u < {\alpha _2} < \tilde \alpha _2^u.
\vspace{-0.2cm}
\end{equation}
As a result, when the following conditions are satisfied with $a,b,c$ defined in (\ref{dsef}):
\begin{equation}\label{jpojp}
 {\rm{ C6 }}:{\rm{ C_X, }}\ ab - c < 0{\rm{ \ and\  }}{\left( {\frac{c}{{ab}}} \right)^{1/(b - 1)}}\left( {1 - b} \right){\rm{ + }}b < 0,
\end{equation}
the average power required for RRH 1 and RRH 2, denoted as $T_{{\rm{RR}}{{\rm{H}}_1}}^{{\rm{C6}}}$ and $T_{{\rm{RR}}{{\rm{H}}_2}}^{{\rm{C6}}}$ respectively, is similar to the expressions of $T_{{\rm{RR}}{{\rm{H}}_1}}^{{\rm{C5}}}$ and $T_{{\rm{RR}}{{\rm{H}}_2}}^{{\rm{C5}}}$ except that the integration interval of $\alpha _2$ becomes (\ref{oihe}).

\emph{Case II:} If the following condition is satisfied:
 \begin{equation}\label{dokpof}
 \frac{{{\lambda _1}}}{{{\lambda _2}}}{\alpha _2} < A,
\end{equation}
the feasible region of $\alpha _1$ is $B>\alpha _1>A$, where $B$ is defined in (\ref{frsf}).  The above condition is equivalent to $h(\alpha _2)<0$ with $a,b,c$ defined in (\ref{dsef}). To guarantee the existence of a positive  $\alpha _2$, the third condition in Lemma 3 should be satisfied. Assuming that this condition is satisfied, the feasible region of $\alpha _2$ is given by $ \alpha _2^l < {\alpha _2} <  \alpha _2^u$, where ${ \alpha _2^l}$ and ${ \alpha _2^u}$ are the solutions of $h(\alpha _2)=0$ with $a,b,c$ defined in (\ref{dsef}). Again, by using the relations in (\ref{relation}) and the condition (\ref{dewaa}), the feasible region of $\alpha _2$ is given by $ \alpha _2^l < {\alpha _2} < \alpha _2^u$.

Hence, if the following conditions are satisfied with $a,b,c$ defined in (\ref{dsef}):
\begin{equation}\label{jpojpsade}
 {\rm{ C7 }}:{\rm{ C_X, }}\ ab - c < 0{\rm{ \ and\  }}{\left( {\frac{c}{{ab}}} \right)^{1/(b - 1)}}\left( {1 - b} \right){\rm{ + }}b < 0,
\end{equation}
that the average powers required for RRH 1 and RRH 2 are respectively given by
\begin{align}
&T_{{\rm{RR}}{{\rm{H}}_1}}^{{\rm{C7}}} =\nonumber \\
& \frac{{P_{\rm{1}}^{{\rm{peak}}}}}{{\Gamma \left( m \right)}}\int_{ \alpha _2^l}^{ \alpha _2^u} {\left[ {\gamma \left( {m,\frac{{mB}}{{{{\bar \alpha }_1}}}} \right) - \gamma \left( {m,\frac{{mA}}{{{{\bar \alpha }_1}}}} \right)} \right]f({\alpha _2})} d{\alpha _2},\\
&T_{{\rm{RR}}{{\rm{H}}_2}}^{{\rm{C7}}} = {{\tilde J}_1} - {{\tilde J}_2},
\end{align}
where ${{\tilde J}_1}$ and ${{\tilde J}_2}$ are given by
\[\begin{array}{l}
{{\tilde J}_1} = \frac{1}{{\Gamma \left( m \right)}}\int_{ \alpha _2^l}^{ \alpha _2^u} C \left( {\gamma \left( {m,\frac{{mB}}{{{{\bar \alpha }_1}}}} \right) - \gamma \left( {m,\frac{{mA}}{{{{\bar \alpha }_1}}}} \right)} \right)f\left( {{\alpha _2}} \right)d{\alpha _2},\\
{{\tilde J}_2} = \frac{{{{\bar \alpha }_1}P_1^{{\rm{peak}}}}}{{m\Gamma \left( m \right)}}\int_{\alpha _2^l}^{\alpha _2^u} {\frac{1}{{{\alpha _2}}}\left( {H - L} \right)f\left( {{\alpha _2}} \right)} d{\alpha _2}.
\end{array}\]
where $H$ and $L$ are respectively given by
\[H = \gamma \left( {m + 1,\frac{{mB}}{{{{\bar \alpha }_1}}}} \right),L = \gamma \left( {m + 1,\frac{{mA}}{{{{\bar \alpha }_1}}}} \right).\]

\vspace{-0.5cm}
\subsection{Discussion of the results}
\vspace{-0.1cm}
\begin{table}[!t]
\renewcommand{\arraystretch}{1.1}
\caption{Main results when $I=2$, and $\frac{\lambda_1}{{\alpha}_1}\le\frac{\lambda_2}{{\alpha}_2}$}
\vspace{-0.3cm}
\label{tab1}
\centering
\begin{tabular}{|c|c|c|}
\hline
&\textbf{Conditions}   & \textbf{Average Power} \\
\hline
 \multirow{4}{*}{$\left| {{\cal I}'} \right|=1$}&C1 &  $T_{{\rm{RRH}}_1}^{{\rm{C1}}}$ \\
\cline{2-3}
 &C2 &  $T_{{\rm{RRH}}_1}^{{\rm{C2}}}$ \\
\cline{2-3}
&C3&  $T_{{\rm{RRH}}_1}^{{\rm{C3}}}$\\
\hline
\multirow{4}{*}{$\left| {{\cal I}'} \right|=2$}&C4 & $T_{{\rm{RRH}}_1}^{{\rm{C4}}},T_{{\rm{RRH}}_2}^{{\rm{C4}}}$\\
\cline{2-3}
&C5 &  $T_{{\rm{RRH}}_1}^{{\rm{C5}}},T_{{\rm{RRH}}_2}^{{\rm{C5}}}$\\
\cline{2-3}
&C6 &  $T_{{\rm{RRH}}_1}^{{\rm{C6}}},T_{{\rm{RRH}}_2}^{{\rm{C6}}}$\\
\cline{2-3}
&C7 &  $T_{{\rm{RRH}}_1}^{{\rm{C7}}},T_{{\rm{RRH}}_2}^{{\rm{C7}}}$\\
\hline
\end{tabular}
\vspace{-0.5cm}
\end{table}

In this subsection, we summarize the results discussed in the above two subsections in Table \ref{tab1}.  The average power required for each RRH contributed under condition (\ref{paixu}) is given by
\begin{equation}\label{dejoij}
  P_{{\rm{RR}}{{\rm{H}}_1}}^{\frac{{{\lambda _1}}}{{{\alpha _1}}} \le \frac{{{\lambda _2}}}{{{\alpha _2}}}} = \sum\limits_{i = 1}^7 {T_{{\rm{RR}}{{\rm{H}}_1}}^{{\rm{C}}i}\varepsilon ({\rm{C}}i)} ,P_{{\rm{RR}}{{\rm{H}}_2}}^{\frac{{{\lambda _1}}}{{{\alpha _1}}} \le \frac{{{\lambda _2}}}{{{\alpha _2}}}} = \sum\limits_{i = 4}^7 {T_{{\rm{RR}}{{\rm{H}}_2}}^{{\rm{C}}i}\varepsilon ({\rm{C}}i)},
\end{equation}
where $\varepsilon \left( \cdot \right)$ is an indicator function,  defined as
\begin{equation}\label{indifunc}
  \varepsilon \left( {\rm{C}}i \right) = \left\{ {\begin{array}{*{20}{l}}
{1,\;{\rm{if}}\;{\rm{Condition}}\  {\rm{C}}i\  {\rm{holds}},}\\
{0,\;{\rm{otherwise}}.}
\end{array}} \right.
\end{equation}

Then, the average power required for each RRH is given by
\begin{equation}\label{FRR}
{P_{{\rm{RR}}{{\rm{H}}_1}}} = P_{{\rm{RR}}{{\rm{H}}_1}}^{\frac{{{\lambda _1}}}{{{\alpha _1}}} \le \frac{{{\lambda _2}}}{{{\alpha _2}}}} + P_{{\rm{RR}}{{\rm{H}}_1}}^{\frac{{{\lambda _1}}}{{{\alpha _1}}} > \frac{{{\lambda _2}}}{{{\alpha _2}}}},{P_{{\rm{RR}}{{\rm{H}}_2}}} = P_{{\rm{RR}}{{\rm{H}}_2}}^{\frac{{{\lambda _1}}}{{{\alpha _1}}} \le \frac{{{\lambda _2}}}{{{\alpha _2}}}} + P_{{\rm{RR}}{{\rm{H}}_2}}^{\frac{{{\lambda _1}}}{{{\alpha _1}}} > \frac{{{\lambda _2}}}{{{\alpha _2}}}},
\end{equation}
where $P_{{\rm{RR}}{{\rm{H}}_1}}^{\frac{{{\lambda _1}}}{{{\alpha _1}}} > \frac{{{\lambda _2}}}{{{\alpha _2}}}}$ and $P_{{\rm{RR}}{{\rm{H}}_2}}^{\frac{{{\lambda _1}}}{{{\alpha _1}}} > \frac{{{\lambda _2}}}{{{\alpha _2}}}}$ denotes the average transmit power under condition ${\frac{{{\lambda _1}}}{{{\alpha _1}}} > \frac{{{\lambda _2}}}{{{\alpha _2}}}}$ for RRH 1 and RRH 2, which can be calculated as the condition of ${\frac{{{\lambda _1}}}{{{\alpha _1}}} \le \frac{{{\lambda _2}}}{{{\alpha _2}}}}$.

\begin{figure*}
\centering
\includegraphics[width=\linewidth]{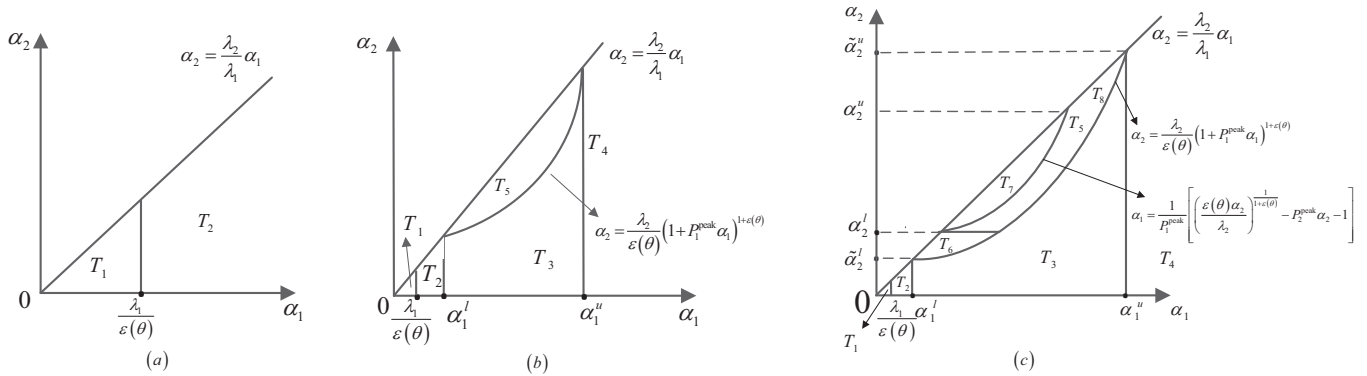}
\vspace{-0.4cm}
\caption{The partitions of the region of ${\frac{{{\lambda _1}}}{{{\alpha _1}}} > \frac{{{\lambda _2}}}{{{\alpha _2}}}}$ under different dual variables: (a) $\lambda_1=\lambda_2=1$; (b) $\lambda_1=1/5, \lambda_2=1$; (c) $\lambda_1=\lambda_2=1/10$. The system parameters are set as \cite{cheng2016}: Time frame of length $T_f=0.2\  {\rm{ms}}$, the system bandwidth $B=100\  {\rm{KHz}}$, delay QoS exponent $\theta=({\rm{ln}}2)/20$ so that $\varepsilon(\theta)=1$, peak power constraint  $P_i^{\rm{peak}}=1\  {\rm{W}}, \forall i=1,2$. }
\label{casespec}
\vspace{-0.2cm}
\end{figure*}

To provide more insights concerning the joint power allocation results for the case of $I=2$, in Fig.~\ref{casespec} we plot the regions corresponding to different cases of the dual variables. For clarity, we only consider the case of  ${\frac{{{\lambda _1}}}{{{\alpha _1}}} > \frac{{{\lambda _2}}}{{{\alpha _2}}}}$.

Fig.~\ref{casespec}-(a) corresponds to the case of $\lambda_1=\lambda_2=1$. From this figure, we can see that our proposed joint power allocation algorithm divides the region of ${\frac{{{\lambda _1}}}{{{\alpha _1}}} > \frac{{{\lambda _2}}}{{{\alpha _2}}}}$ into two exclusive regions by the solid lines. If $(\alpha_1,\alpha_2)$ falls into the region $T_1$, both RRHs remian silent. On the other hand, if $(\alpha_1,\alpha_2)$ falls into the region $T_2$, Condition C2 is satisfied and only RRH 1 will transmit with a non-zero power, but less than the peak power.

Fig.~\ref{casespec}-(b) corresponds to the case of $\lambda_1=1/5, \lambda_2=1$. In this case, the region of ${\frac{{{\lambda _1}}}{{{\alpha _1}}} > \frac{{{\lambda _2}}}{{{\alpha _2}}}}$ is divided into five exclusive regions. Similarly, in region $T_1$, none of the RRHs transmit. In region $T_2$ and $T_4$, Condition C3 is satisfied and only RRH 1 is assigned non-zero power for data transmission, but less than the peak power. If $(\alpha_1,\alpha_2)$ falls into region $T_3$, Condition C1 is satisfied, and RRH 1 will transmit at peak power and RRH 2 still remains silent. However, if $(\alpha_1,\alpha_2)$ falls into the region $T_5$, Condition C5 holds, RRH 1 will transmit at peak power and RRH 2 is allocated positive power that is lower than the peak power.

Fig.~\ref{casespec}-(c) corresponds to the case of $\lambda_1=\lambda_2=1/10$. In this case, the region is partitioned into as many as eight regions. Regions $T_1$-$T_4$ are the same as the case in Fig.~\ref{casespec}-(b). If $(\alpha_1,\alpha_2)$ falls into regions $T_6$ and $T_8$, Conditions C6 is satisfied, hence RRH 1 will transmit at its peak power and RRH 2 is assigned non-zero power for its transmission, but its power is less than the peak power. If $(\alpha_1,\alpha_2)$ falls into region $T_7$, Condition C7 is satisfied. Similarly, RRH 1 transmits at its maximum power and RRH 2 transmits at a non-zero power below its peak power. On the other hand, if $(\alpha_1,\alpha_2)$ falls into region $T_7$, Condition C4 is satisfied and both RRHs will transmit at their peak power.

It is interesting to find that with the reduction of the dual variables, more RRHs will transmit at non-zero power or even the peak power.  This can be explained as follows. The dual variables can be regarded as a pricing factor, where a lower dual variable will encourage the RRHs to be involved in transmission due to the low cost.

\vspace{-0.2cm}
\section{Power Allocation in Extreme Cases}\label{extremeajeoioif}
In this section, we discuss the power allocation for two extreme cases: 1) when $\theta\rightarrow 0$, representing no delay requirement; 2) when $\theta\rightarrow \infty$ and $P_i^{\rm{peak}}\rightarrow \infty, \forall i\in \cal I$, representing the extremely strict zero delay requirement.
\subsection{$\theta\rightarrow 0$ (No delay requirements)}

When $\theta\rightarrow 0$, the system is delay-tolerant. The EC maximization problem is equivalent to the ergodic capacity maximization problem, which is given by
\begin{equation}
\label{turnedeta-pro}
\begin{array}{l}
\mathop {\max }\limits_{\left\{ {{p_i}\left( {\bm{\alpha }} \right),\forall i} \right\}} \;\;{\mathbb{E}_{\bm{\alpha }}}\left[ {{{\log }_2}\left( {1 + \sum\nolimits_{i \in {\cal I}} {{p_i}\left( {\bm{\alpha }} \right){\alpha _i}} } \right)} \right] \\
\;\;{\rm{s}}{\rm{.t}}{\rm{.}}\;\;\;\;{\mathbb{E}_{\bm{\alpha }}}\left[ {{p_i}\left( {\bm{\alpha }} \right)} \right] \le P_i^{\text{avg}},\;\;\forall i\\
\qquad\quad 0\leq {p_i}\left( {\bm{\alpha }} \right) \le P_i^{\text{peak}}, \;\;\forall i.
\end{array}
\end{equation}
By using a similar method as in Section \ref{genralcase}, the optimal transmit power at each RAU is given by
\vspace{-0.2cm}
\begin{equation}
\label{powerallocation}
p_{\pi \left( a \right)}^* = \left\{ \begin{array}{l}
P_{\pi \left( a \right)}^{{\rm{peak}}},\qquad\qquad\qquad\quad a < \left| {{\cal I}'} \right|;\\
\min \left\{ {P_{\pi \left( {\left| {{\cal I}'} \right|} \right)}^{{\rm{peak}}},{{\tilde T}_{\pi (a)}}} \right\},\;a = \left| {{\cal I}'} \right|;\\
0,\qquad\qquad\qquad\qquad\quad a > \left| {{\cal I}'} \right|;
\end{array} \right.
\end{equation}
where ${{\tilde T}_{\pi (a)}}$ is given by
\[{{\tilde T}_{\pi (a)}} = \frac{1}{{{\alpha _{\pi \left( {\left| {{{\cal I}^\prime }} \right|} \right)}}}}\left( {\frac{1}{{\ln 2}}\cdot\frac{{{\alpha _{\pi \left( {\left| {{{\cal I}^\prime }} \right|} \right)}}}}{{{\lambda _{\pi \left( {\left| {{{\cal I}^\prime }} \right|} \right)}}}} - \sum\limits_{a = 1}^{\left| {{{\cal I}^\prime }} \right| - 1} {P_{\pi \left( a \right)}^{{\rm{peak}}}} {\alpha _{\pi \left( a \right)}} - 1} \right)\]
with $\left| {{\cal I}'} \right|$ being the largest value of $x$ such  that
\begin{equation}
\label{largestIcondition}
\frac{1}{{{\rm{In}}2}} \cdot \frac{{{\alpha _{\pi (x)}}}}{{ {\lambda _{\pi (x)}}}} > 1 + \sum\limits_{a = 1}^{x - 1} {P_{\pi \left( a \right)}^{{\rm{peak}}}{\alpha _{\pi (a)}}}
\end{equation}
and $\pi$ is a permutation over $\mathcal{I}$ so that $\frac{\lambda_{\pi(i)}}{\alpha_{\pi(i)}}\le\frac{\lambda_{\pi(j)}}{\alpha_{\pi(j)}}$, when $i<j, i,j\in \cal I$.

When there is only one RRH, the optimal power allocation reduces to the conventional water-filling solution that is given by
\begin{equation}\label{frhiohtfeaf}
 p_1^* = \left[ {\frac{1}{{{\lambda _1}\ln 2}} - \frac{1}{{{\alpha _1}}}} \right]_0^{P_1^{{\rm{peak}}}},
\end{equation}
where $\left[ x \right]_0^y$ represents that the value is zero if $x<0$, $x$ if $0<x<y$, and  $y$ if $x>y$.

\vspace{-0.4cm}
\subsection{$\theta\rightarrow \infty$ and $P_i^{\rm{peak}}\rightarrow \infty, \forall i\in \cal I$ (Extremely strict delay requirement)}
For this case, the system cannot tolerate any delay, and the EC is equal to the zero-outage capacity. Hence, the optimal power allocation for each RRH reduces to the channel inversion associated with a fixed data rate \cite{Goldsmith19972}
given by
\begin{equation}\label{rfhightu}
  p_i^* = \frac{{{\beta _i}}}{{{\alpha _i}}},\forall i\in {\cal I},
\end{equation}
where $\beta _i$ is specifically selected to satisfy the average power constraints. For the Nakagami-$m$ channel defined in (\ref{rheisghu}), we have
\begin{equation}\label{fefref}
  {{\mathbb{E}}_{{\alpha _i}}}\left\{ {\frac{1}{{{\alpha _i}}}} \right\} = \left\{ \begin{array}{l}
\frac{m}{{\left( {m - 1} \right){{\bar \alpha }_i}}},m > 1\\
0,\qquad\  1 \geq m.
\end{array} \right.
\end{equation}
Then, $\beta _i$ can be calculated as
\begin{equation}\label{defaeraf}
 {\beta _i} = \left\{ \begin{array}{l}
\frac{{\left( {m - 1} \right){{\bar \alpha }_i}P_i^{{\rm{avg}}}}}{m},m > 1\\
0,\qquad\qquad 1 \geq m.
\end{array} \right.
\end{equation}
The EC can be obtained as
\begin{equation}\label{feaf}
{\rm{EC}} = \left\{ \begin{array}{l}
{T_f}B{\log _2}(1 + \sum\nolimits_{i \in \cal I} {{\beta _i}} ),m > 1\\
0,\qquad \qquad \qquad\qquad\ \  1 \ge m.
\end{array} \right.
\end{equation}
Note that when $m=1$, the EC is equal to zero, which is consistent with the result in \cite{Goldsmith19972}, namely that the zero-outage capacity for Rayleigh fading is equal to zero.  However, when $m>1$, the EC increases with $m$. This coincides with the intuition that with the increase of $m$, the channel variations reduce and the channel becomes more stable. As a result, the queue-length variations are reduced and hence higher delay requirements can be satisfied.

\section{Extension to the Multiuser Case}\label{hfehreg}

The above sections focus on the single user case. However, in C-RAN, due to the powerful computational capability of the BBU pool, C-RAN is designed with the goal of serving multiple users. To simply the analysis, we assume that the RRHs transmit the signals to different users in orthogonal channels to avoid multiuser interference. Similar assumptions have been made in \cite{Stephen2017,liangliu2015,chunlonghejsac,fheith}.

Let us denote the total number of users by $K$  and ${\cal K} = \left\{ {1, \cdots ,K} \right\}$. The QoS exponent for user $k$ is denoted as $\theta_k$, and the set of QoS exponents is denoted as $\bm\theta=\{\theta_1,\cdots,\theta_K\}$. The channel's power gain from RRH $i$ to user $k$ is denoted as $\alpha_{i,k}$. The set of channel gains from all RRHs to user $k$ is denoted as ${\bm{\alpha} _k} = \left\{ {{\alpha _{1,k}}, \cdots ,{\alpha _{I,k}}} \right\}$. The set of all channel gains in the network is denoted as ${\bm\alpha}=\{{\bm\alpha}_1, \cdots, {\bm\alpha}_K\}$ . For simplicity, the peak power constraints are not considered.  We aim for optimizing the power allocation to maximize the sum EC of all users subject to the per-RRH average power constraints. This optimization problem is formulated as
\begin{subequations}\label{ifdfreafro}
\begin{align}
\mathop {\max }\limits_{{\left\{ {p_{i,k}(\bm\nu)},\forall i,k\right\}}}\;\;\;& \sum\limits_{k \in {\cal K}} { - \frac{1}{{{T_fB\theta _k}}}\log \left( {{\mathbb{E}_{{\bm\alpha}}}\left[ {{e^{ - {\theta _k}{R_k}(\bm\nu )  }}} \right]} \right)} \\
{\rm{s.t.}}\;\;\;&  {\mathbb{E}_{\bm\alpha} }\left[ {\sum\limits_{k \in {\cal K}} {{p_{i,k}}(\bm\nu )} } \right]\le P^{\text{avg}}_i,\forall i \in \mathcal{I},\\
&0\le{p_{i,k}(\bm\nu)},\forall i \in \mathcal{I}, k\in \cal K,
\end{align}
\end{subequations}
where  $\bm\nu=\{\bm\alpha, \bm\theta\}$ represents the network condition that includes both the channel's power gains and the QoS exponent requirements, ${R_k}(\bm\nu )$ is the data rate achieved by user $k$ that is given by ${R_k}(\bm\nu ) = {T_f}B{{\log }_2}\left( {1 + \sum\nolimits_{i \in {\cal I}} {{p_{i,k}}(\bm\nu ){\alpha _{i,k}}} } \right)$.

Note that in the multiuser case, each user's power allocation depends both on its own channel condition and on its own delay requirement as well as those of other users'. Furthermore, the power allocation solution of each user is coupled with the per-RRH average power constraints, which complicates the analysis. Unlike in the single user case, Problem (\ref{ifdfreafro}) cannot be decomposed into several independent subproblems for each channel fading state. The reason for this is that the objective function in Problem (\ref{ifdfreafro}) cannot be transformed into the format of Problem (\ref{mainproblem}). However, since the EC is a concave function \cite{Helmy13} and the constraints are linear, Problem (\ref{ifdfreafro}) still remains a convex optimization problem. Hence, the Lagrangian duality method can still be used to solve this problem.

Specifically, we introduce the dual variables of $\lambda_i, \forall i$, and $\delta_{i,k}, \forall i, k$ for the corresponding constraints in Problem (\ref{ifdfreafro}). The KKT conditions for the optimal solutions of Problem (\ref{ifdfreafro}) are given by
\begin{subequations}
\begin{align}
- \frac{{\varepsilon ({\theta _k}){\alpha _{i,k}}{{{Z_k}(\bm\nu )}^{ \!-\! \varepsilon ({\theta _k}) \!-\! 1}}}}{{\kappa _k}}f(\bm\alpha ) \!+\! {\lambda _i}f(\bm\alpha )\! -\! {\delta _{i,k}} \!= 0, \forall i\label{kfeafefp}\\
{\lambda _i}\left( {{{\mathbb{E}}_{\bm\alpha} }\left[ {\sum\nolimits_{k \in \cal K} {{p_{i,k}^*}(\bm\nu )} } \right] - P_i^{{\rm{avg}}}} \right) = 0, \forall i\label{kkawqqdu}\\
\delta_{i,k}^*p_{i,k}^*(\bm\nu )=0, \forall i,k\label{kkrfta}\\
p_{i,k}^*(\bm\nu ) \ge 0, \forall i,k\label{kktEFsactive}
\end{align}
\end{subequations}
where $f(\bm\alpha)$ represents the joint PDF of $\bm\alpha$, ${Z_k}(\bm\nu ) = 1 + \sum\nolimits_{i \in \cal I} {p_{i,k}^*(\bm\nu ){\alpha _{i,k}}}$ and $\varepsilon ({\theta _k}) = {{{\theta _k}{T_f}B} \mathord{\left/
 {\vphantom {{{\theta _k}{T_f}B} {\ln 2}}} \right.
 \kern-\nulldelimiterspace} {\ln 2}}$, ${\kappa _k}$ is given by
 \begin{equation}\label{reagr}
  \kappa _k=T_fB{{\theta _k}{{\mathbb{E}}_{\bm\alpha} }\left[ {{{{Z_k}(\bm\nu )}^{ - \varepsilon ({\theta _k})}}} \right]}.
 \end{equation}
To satisfy the above KKT conditions, we first find the optimal power allocation associated with the fixed dual variables $\lambda_i,\forall i$, then update the dual variables by using the sub-gradient method.

Given $\lambda_i,\forall i$, we then have the following theorem.

\textbf{Theorem 3}: There is at most one RRH serving each user and the index of this RRH is given by
\begin{equation}\label{adfefrea}
 {i^*} = \mathop {\arg \min }\limits_{i \in {\cal I}} \frac{{{\lambda _i}}}{{{\alpha _{i,k}}}},
\end{equation}
while the corresponding power allocation is given by
\begin{equation}\label{fraer}
 p_{i^*,k}^* = {\left[ {{{\left( {\frac{{{\varepsilon ({\theta _k})}}}{{\kappa _k{\lambda _{i^*}}}}} \right)}^{\frac{1}{{1 + \varepsilon \left( {{\theta _k}} \right)}}}}\alpha _{i^*,k}^{ - \frac{{\varepsilon \left( {{\theta _k}} \right)}}{{1 + \varepsilon \left( {{\theta _k}} \right)}}} - \frac{1}{{{\alpha _{i^*,k}}}}} \right]^ + },
\end{equation}
where ${\left[ x \right]^ + } = \max \left\{ {0,x} \right\}$.

\emph{\textbf{Proof}}: The proof is similar to that of the single user scenario of Section \ref{genralcase}, which is omitted for simplicity. \hfill\rule{2.7mm}{2.7mm}

Although each user is served by only one RRH for each channel state, they still benefit from the multi-RRH diversity, as seen in (\ref{adfefrea}). Similarly, the online calculation method used for the single user case can be adopted to approximate the value of  $\kappa _k$.

\vspace{-0.2cm}
\section{Simulations}\label{simulation}

In this section, we characterize the performance of our proposed algorithm. We consider a C-RAN network covering a square area of 2 km $\times$ 2 km. The user is located at the center of this region and the RRHs are independently and uniformly allocated in this area. Unless otherwise stated, we adopt the same simulation parameters as in \cite{cheng2016}: Time frame of length $T_f=0.1\  {\rm{ms}}$;  system bandwidth of $B=200\  {\rm{KHz}}$; average power constraint of $P_i^{\rm{avg}}=0.5\  {\rm{W}},\forall i$; peak power constraint of $P_i^{\rm{peak}}=1\  {\rm{W}}, \forall i$; QoS exponent of $\theta=0.05$; Nakagami fading parameter $m=2$. The path-loss is modeled as $PL_{i} \!=\! 148.1 \!+\! 37.6{\log _{10}}d_{i}\ ({\rm{dB}})$ \cite{access2010further}, where $d_{i}$ is the distance between the $i$th RRH and the user measured in km. The channel also includes the log-normal shadowing fading with zero mean and 8 dB standard derivation. The noise density power is -174 dBm/Hz \cite{access2010further}.

We compare our algorithm to the following algorithms:
\begin{enumerate}
  \item \textit{Nearest RRH serving algorithm}: In this algorithm, the user is only served by its nearest RRH, and the optimization method proposed in \cite{cheng2016} for point-to-point systems is adopted to solve the power allocation problem in this setting. This algorithm is proposed to show the benefits of cooperative transmission in C-RAN.
  \item \textit{Constant power allocation algorithm}: In this algorithm, the transmit power is set to be equal to the average power constraint for any time slots. Since the peak power limit is higher than the average power limit, both the average power constraints and peak power constraints can be satisfied. This approach was assumed to illustrate the benefits of dynamic power allocation proposed in this paper.
  \item \textit{Independent power allocation algorithm}: In this algorithm, each RRH $i$ independently optimizes its power allocation without considering the joint channel conditions of the other RRHs. The optimization problem is given by
      \vspace{-0.1cm}
      \begin{subequations}\label{iidne}
           \begin{align}
              \mathop {\max }\limits_{{{p_i(\bm\nu)}}}\;\;\;&  - \frac{1}{\theta}{\log}\left(\mathbb{E}_{\bm \alpha} {\left[ {{e^{ - \theta T_fB \log_2\left( {1 +  {{p_i}(\bm\nu){\alpha_i}} } \right)}}} \right]} \right)\\
              {\rm{s.t.}}\;\;\;&  \mathbb{E}_{\bm\alpha}\left[{{p_i(\bm\nu)}} \right] \le P^{\text{avg}}_i,\\
              &0\le{p_i(\bm\nu)} \le {P_i^{\text{peak}}}.
           \end{align}
      \end{subequations}
      The optimal power allocation solution is given by (\ref{hideo}). This algorithm is invoked for showing the benefits of jointly optimizing the power allocation according to the joint channel conditions of all RRHs.
\item \textit{Ergodic capacity maximization algorithm}: This algorithm aims for maximizing the ergodic capacity without considering the delay requirement. The optimal power allocation solution is given by (\ref{powerallocation}).
\item \textit{Channel inversion algorithm}: This algorithm imposes an extremely strict delay requirement. The power allocation is the channel inversion given by (\ref{rfhightu}).
\end{enumerate}

In the following, we first consider the case of $I=2$, where the average power required for each RRH can be numerically obtained according to the results of Section \ref{specialcaseoftwo}. Then, we study the more general case associated with more than two RRHs.
\begin{figure}
\centering
\includegraphics[width=\linewidth]{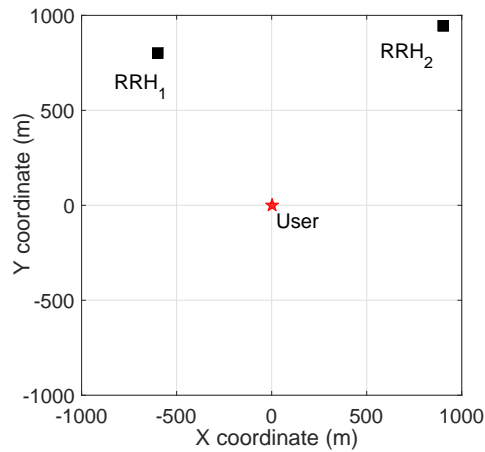}
\caption{Random network topology with two RRHs. The user is located at the center of the C-RAN. The coordinates of RRH 1 and RRH 2 are given by $\left[ {{\rm{ - }}600,800} \right]$ and $\left[ {900,946} \right]$, respectively. The CPNRs are $\bar \alpha_1=3.89$ and $\bar \alpha_2=1.43$.}
\label{scenario1}
\end{figure}
\begin{figure}
\centering
\includegraphics[width=\linewidth]{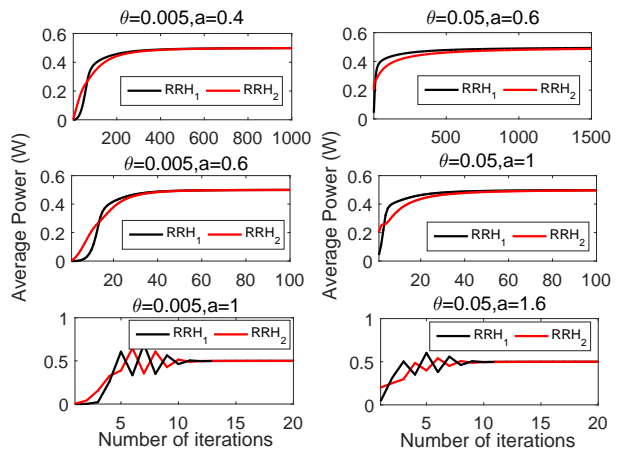}
\caption{Convergence behaviour of the proposed algorithm for the special case of two RRHs under different step parameters. Left three subplots correspond to the QoS exponent $\theta=0.005$, and the right three correspond to $\theta=0.05$.}
\label{convergentwo}
\end{figure}

\begin{figure}
\centering
\includegraphics[width=\linewidth]{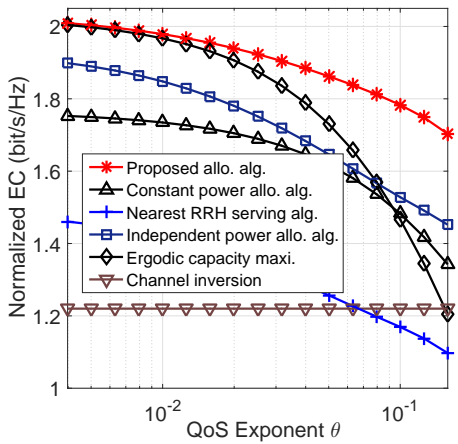}
\caption{Normalized EC for various algorithms vs QoS exponent $\theta$.}
\label{impactQoS}
\end{figure}

\begin{figure}
\centering
\includegraphics[width=\linewidth]{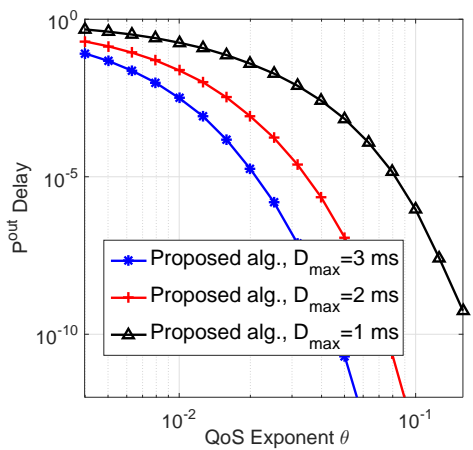}
\caption{Delay-outage probability versus QoS exponent $\theta$ for various values of $D_{\rm{max}}$ for our proposed algorithm.}
\label{outageprobality}
\end{figure}

Fig.~\ref{scenario1} shows a randomly generated network topology relying on two RRHs. The average CPNRs related to RRH 1 and RRH 2 are then given by $\bar \alpha_1=3.89$ and $\bar \alpha_2=1.43$. Figs.~\ref{convergentwo}-\ref{outageprobality} are based on this network topology. In this setting, the user is only served by RRH 1 due to its shorter distance compared to RRH 2, when using the nearest RRH algorithm.

In Fig.~\ref{convergentwo}, we study the convergence behaviour of the proposed algorithm under different delay exponents and step-size parameters. The left three subplots correspond to the QoS exponent $\theta=0.005$, while the right three correspond to $\theta=0.05$. It can be observed from this figure that the convergence speed mainly depends on step-size parameter $a$. For both cases of $\theta$, a smaller step-size parameter leads to slower convergence and larger one leads to faster convergence speed. For example, for the case of $\theta=0.005$, nearly 700 iterations are required for convergence when $a=0.4$, while only 20 iterations are needed when $a=1$. However, since the subgradient method is not an ascent method, a larger value of $a$ may lead to a fluctuation of the power value. A small value of $a$ yields a smooth but slow convergence. Hence, the step-size parameter should be carefully chosen to strike a tradeoff between the convergence speed and the smoothness of the power curve.  It should be emphasized that the main advantage for the case of two RRHs is that the Lagrange dual variables can be calculated in an off-line manner by using the results of Section \ref{specialcaseoftwo}, and stored in the memory.

 Let us now study the impact of the delay requirements on the EC performance for various algorithms in Fig.~\ref{impactQoS}. The normalized EC performance (which is defined as the EC divided by $B$ and $T_f$ with the unit of `bit/s/Hz') is considered. As expected, when the delay-QoS requirement becomes more stringent, i.e., the value of QoS exponent $\theta$ increases, the normalized EC achieved by all algorithms is reduced. By employing two RRHs for jointly serving the user to exploit higher spatial degrees of freedom,  the proposed algorithm significantly outperforms the nearest RRH based algorithm, where only one RRH is invoked for transmission. For example, for $\theta=0.1$ the normalized EC provided by the former algorithm is roughly 50\% higher than that of the latter algorithm. It is observed that the performance of the latter algorithm is even inferior to that of the naive constant power allocation algorithm. Since our proposed algorithm adapts its power allocation to delay-QoS requirement, whilst additionally taking into account the channel conditions, it achieves a higher normalized EC than the constant power allocation algorithm, and the performance gain attained is about 20\% when $\theta=0.1$. By optimizing the power allocation based on the joint channel conditions, the proposed algorithm performs better than the independent one, where the joint relationship of the channel conditions is ignored. The performance gain is more prominent when the delay-QoS requirement is very stringent, which can be up to 18.6\% in this example. The proposed algorithm has similar performance to that of the ergodic capacity maximization  algorithm for small  $\theta$, while the performance gain provided by the proposed algorithm becomes higher for a large $\theta$. For a high $\theta$, the normalized EC achieved by the ergodic capacity maximization  algorithm is even lower than that of the constant power allocation algorithm. It is seen from Fig.\ref{impactQoS} that the performance of the channel inversion based power allocation method is much worse than that of our proposed algorithm, since it aims for an extremely strict delay budget rather than adaptively adjusting the power allocation according to the delay requirements.

Fig.~\ref{outageprobality} shows the delay-outage probability versus QoS exponent $\theta$ for various values of $D_{\rm{max}}$ for our proposed algorithm. In this example, the UL and DL transmission duration and the fronthaul delay are set as ${D_T}={D_F}=T_f=0.1$ ms \cite{changyang2018twc}. Then, the corresponding queueing delay is given by $D_q=D_{\rm{max}}-{D_T}-{D_F}=D_{\rm{max}}-0.2$ ms. The maximum incoming data rate is set to be equal to the EC, i.e. to $\mu  = {\rm{EC}}(\theta )$. The method in \cite{Dapeng2003} is used for calculating the delay-outage probability. As expected, the delay outage probability increases with the reduction of $D_{\rm{max}}$. By appropriately choosing the QoS exponent $\theta$, the delay outage probability can be reduced below $10^{-8}$. For example, for the case of $D_{\rm{max}}=1\ \rm{ms}$, the delay outage probability is as low as $4\times 10^{-10}$ when $\theta=0.1585$, which satisfies the stringent delay requirement of the ultra-reliable low latency communications (URLLC) in 5G \cite{Andrews14}. For a given target $P_{\rm{delay}}^{\rm{out}}$, the corresponding value of $\theta$ obtained from Fig.~\ref{outageprobality} can be used in Fig.~\ref{impactQoS} to find the achievable EC. For example, when the delay outage probability requirement is $P_{\rm{delay}}^{\rm{out}}=10^{-6}$ for the case of $D_{\rm{max}}=1\ \rm{ms}$, the corresponding $\theta$ is given by 0.1. The resultant normalized EC becomes 1.78 bit/s/Hz according to Fig.~\ref{impactQoS}.
\begin{figure}
\centering
\includegraphics[width=\linewidth]{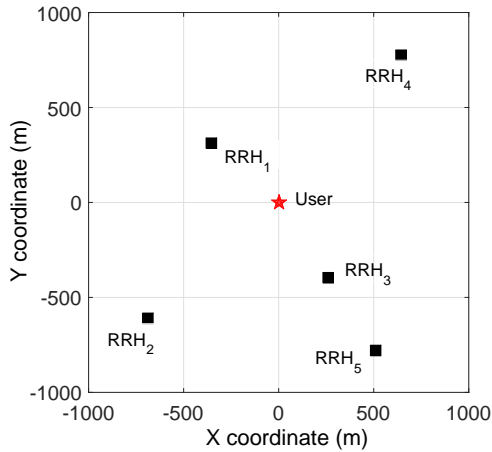}
\caption{Random network topology with five RRHs.}
\label{scenario2}
\end{figure}
\begin{figure}
\centering
\includegraphics[width=\linewidth]{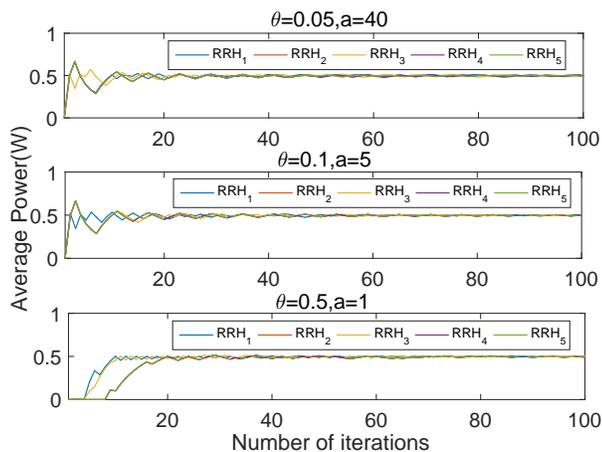}
\caption{Convergence behaviour of the online tracking method.}
\label{convergencgeneral}
\end{figure}
\begin{figure}
\centering
\includegraphics[width=\linewidth]{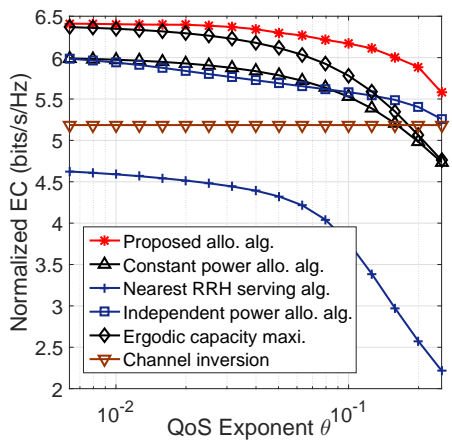}
\caption{Normalized EC for various algorithms vs QoS exponent $\theta$ for the more general case of $I=5$.}
\label{impactdelaygeneralcase}
\end{figure}
We now consider a more general case in Fig.~\ref{scenario2}, where five RRHs are randomly located in a square. The average CPNRs received from these RRHs are then $\bar \alpha_1=64.3$, $\bar \alpha_2=5.3$, $\bar \alpha_3=63.1$, $\bar \alpha_4=3.8$ and $\bar \alpha_5=5.1$.

Based on Fig.~\ref{scenario2}, the convergence behaviour of the online tracking method is shown in Fig.~\ref{convergencgeneral}, where different QoS requirements are tested.  As shown in Fig.~\ref{convergencgeneral}, the online tracking method converges promptly for all the values of $\theta$ considered which is achieved by properly choosing the step parameter $a$. These observations demonstrate the efficiency of the online tracking method. In Fig.~\ref{impactdelaygeneralcase}, we plot the effective capacity performance for various algorithms for this more general scenario. Similar observations can be found from this figure. For example, the normalized EC achieved by all algorithms decrease  with the QoS exponent $\theta$, and the proposed algorithm significantly outperform the other algorithms.

\begin{figure}
\centering
\includegraphics[width=\linewidth]{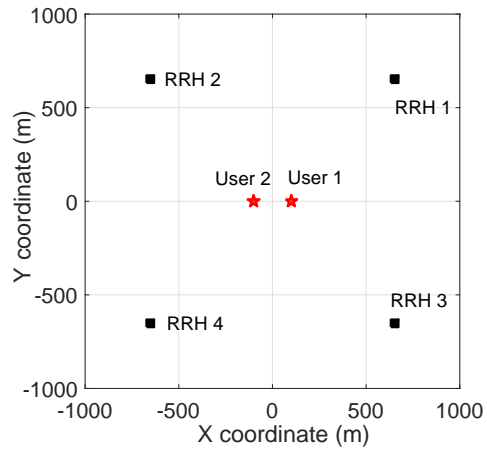}
\caption{Simulated network topology for two users with four RRHs.}
\label{multiuserscebario}
\end{figure}
\begin{figure}
\centering
\includegraphics[width=\linewidth]{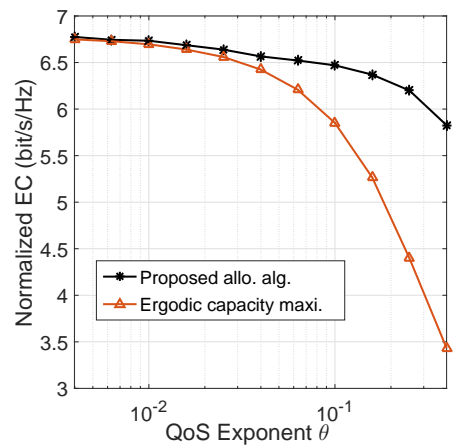}
\caption{The sum normalized EC for various algorithms vs QoS delay exponent $\theta$.}
\label{impfjofkrefjrei}
\end{figure}

Finally, we study the EC performance of the multiuser scenario. The simulation setup is given in Fig.~\ref{multiuserscebario}, where there are two users located at $[-100,0]$ and $[0,100]$, respectively, while the locations of the four RRHs are given by $[650,650]$, $[-650,650]$, $[-650,-650]$, and $[650,-650]$. We should emphasize that our algorithm used for the multiuser scenario is applicable for any network setup. In Fig.~\ref{impfjofkrefjrei}, we compare our proposed EC maximization algorithm of Subsection \ref{hfehreg} to the multiuser ergodic capacity maximization algorithm in terms of the sum EC of the two users. Note that a similar performance trend has been observed to that in Fig.~\ref{impactQoS}. For example, both algorithms have almost the same performance in the low QoS exponent regime, while our proposed EC-oriented algorithm performs much better than the ergodic capacity-oriented algorithm for high exponents. When $\theta=0.4$, the performance gain is up to 2.3 bit/s/Hz.

\vspace{-0.3cm}
\section{Conclusions}\label{conclusion}
We considered joint power allocation for the EC maximization of C-RAN, where the user has to guarantee a specific delay-QoS requirement to declare successful transmission. Both the per-RRH average and peak power constraints were considered. We first showed that the EC maximization problem can be equivalently transformed into a convex optimization problem, which was solved by using the Lagrange dual decomposition method and by studying the KKT conditions. The online tracking method was proposed for calculating the average power for each RRH. For the special case of two RRHs, the expression of average power for each RRH can be obtained in closed-form. We also provided the power allocation solutions for the extreme cases of $\theta\rightarrow \infty$ and $\theta\rightarrow 0$. The simulation results showed that our proposed algorithm converges promptly and performs much better than the existing algorithms, including the conventional ergodic capacity maximization method operating without delay requirements. A delay outage probability of $10^{-9}$ can be achieved at $D_{\rm{max}}=1$ ms by our proposed algorithm through controlling the value of $\theta$, which is very promising for the applications in future URLLC in 5G.

In this paper, we only extend the single-user case to the multiple-user interference-free case. For the more general case where each user suffers from the multi-user interference, the formulated optimization problem is non-convex, where the Lagrangian duality method is not applicable. How to deal with this problem will be left for future work. In addition, the fronthaul capacity is generally limited due to the fast oversampled real-time I/Q digital data streams transmitting over the fronthaul links. How to design the transmission scheme by taking into account this constraint will also be considered in the future work.

\vspace{-0.3cm}
\begin{appendices}
\section{Proof of convexity of Problem (\ref{mainproblem})}\label{convexityproof}
Since the expectation operator is a linear additive operator, we only have to consider the objective function for each new channel fading generation. For simplicity, we omit the dependency of ${p_i}\left( {\bm\nu} \right)$ on ${\bm\nu} $ and  the  objective function can be expressed as
\begin{equation}\label{funy}
 y({\bf{P}}) \buildrel \Delta \over =  {\left( {1 + \sum\nolimits_{i \in {\cal I}} {{p_i}{\alpha _i}} } \right)^{ - \varepsilon (\theta )}},
\end{equation}
where $\bf{P}$ denotes the collection of power allocations. The second partial derivatives of $y({\bf{P}})$ can be calculated as
\begin{equation}\label{ndei}
  \frac{{\partial {y^2}({\bf{P}})}}{{\partial p_i^2}} = \alpha _i^2\varepsilon (\theta )\left( {\varepsilon (\theta ) + 1} \right){\left( {1 + \sum\nolimits_{i \in {\cal I}} {{p_i}{\alpha _i}} } \right)^{ - \varepsilon (\theta ) - 2}}
\end{equation}
and
\begin{equation}\label{def}
  \frac{{\partial {y^2}({\bf{P}})}}{{\partial {p_i}\partial {p_j}}} = {\alpha _i}{\alpha _j}\varepsilon (\theta )\left( {\varepsilon (\theta ) + 1} \right){\left( {1 + \sum\nolimits_{i \in {\cal I}} {{p_i}{\alpha _i}} } \right)^{ - \varepsilon (\theta ) - 2}}.
\end{equation}
Hence, the Hessian matrix of the function $y({\bf{P}})$ is
\[{\nabla ^2}y({\bf{P}}) = \varepsilon (\theta )\left( {\varepsilon (\theta ) + 1} \right){\left( {1 + \sum\nolimits_{i \in {\cal I}} {{p_i}{\alpha _i}} } \right)^{ - \varepsilon (\theta ) - 2}}\bm \alpha {\bm \alpha ^T},\]
which is a positive semidefinite matrix. Obviously, the constraints in Problem (\ref{mainproblem}) are linear. Hence, the proof is completed.\hfill\rule{2.7mm}{2.7mm}

\vspace{-0.4cm}
\section{Proof of lemma 1}{\label{lemma1}}
As $p_i^*>0$, $p_j^*=0$, from (\ref{kkt-u}) and (\ref{kkt-delta}) we have $\mu_i^*\geq0, \delta_i^*=0$ and $\mu_j^*=0, \delta_j^*\geq0$, respectively. Then, from (\ref{kkt-p}) we have
\begin{align}
\mu_i^*=\varepsilon(\theta)(1+\sum\nolimits_{m\in\mathcal{I}}p_m^*\tilde{\gamma}_m)^{-\varepsilon(\theta)-1}
{\alpha}_i-\lambda_i\ge 0,\\
\delta_j^*=-\varepsilon(\theta)(1+\sum\nolimits_{m\in\mathcal{I}}p_m^*{\alpha}_m)^{-\varepsilon(\theta)-1}
{\alpha}_j+\lambda_j \ge 0.
\end{align}
Hence, it follows that
\begin{equation}
\frac{\lambda_i}{{\alpha}_i}\le\varepsilon(\theta)(1+\sum\nolimits_{m\in\mathcal{I}}p_m^*{\alpha}_m)^{-\varepsilon(\theta)-1}\le\frac{\lambda_j}{{\alpha}_j}.
\end{equation}\hfill\rule{2.7mm}{2.7mm}

\vspace{-0.6cm}
\section{Proof of lemma 2}\label{lemma2}
We first prove the first part of Lemma 2. Assume that there are two RRHs $i$ and $j$ that $0<p_i^*<p_i^{\text{peak}}$ and $0<p_j^*<P_j^{\text{peak}}$ for $i\neq j$. Then, according to (\ref{kkt-u}) and (\ref{kkt-delta}), we have $u^*_i=0, \delta^*_i=0$, $u^*_j= 0, \delta^*_j=0$. Substituting them into (\ref{kkt-p}) yields
\begin{align}
-\varepsilon(\theta)(1+\sum\nolimits_{m\in\mathcal{I}}p_m^*{\alpha}_m)^{-\varepsilon(\theta)-1}{\alpha}_i+\lambda_i=0,\\
-\varepsilon(\theta)(1+\sum\nolimits_{m\in\mathcal{I}}p_m^*{\alpha}_m)^{-\varepsilon(\theta)-1}{\alpha}_j+\lambda_j=0.
\end{align}
Hence, we have
\begin{equation}
\frac{\lambda_j}{{\alpha}_j}=\frac{\lambda_i}{{\alpha}_i}.
\end{equation}
Since ${\alpha}_i$ is independent of ${\alpha}_j$, $\lambda_i$ and $\lambda_j$ are fixed, it is concluded that the above equality holds with a zero probability. Thus, there is at most one RRH $i$ with $0<p_i^*<P_i^{\text{peak}}$.

Let us also assume that there are two users $i,k\in\mathcal{I'}$ with $0<p_i^*<P_i^{\text{peak}}$ and $p_k^*=P_k^{\text{peak}}$. By using (\ref{kkt-u}) and (\ref{kkt-delta}), we have $u^*_i=0, \delta^*_i=0$, $u^*_k\geq 0$, and $\delta^*_k=0$. According to (\ref{kkt-p}), it follows that
\begin{equation}
\frac{\lambda_i}{{\alpha}_i}=\varepsilon(\theta)(1+\sum\nolimits_{m\in\mathcal{I}}p_m^*{\alpha}_m)^{-\varepsilon(\theta)-1}\geq\frac{\lambda_k}{{\alpha}_k}.
\end{equation}
Hence, we conclude that $i=\pi(|\mathcal{I'}|)$.\hfill\rule{2.7mm}{2.7mm}

\vspace{-0.4cm}
\section{Proof of theorem 1}\label{theorem1}
 Lemma 2 implies that the optimal solution must be one of the following two cases:
\begin{itemize}
  \item Case I: $p_{\pi(a)}^*=P_a^{\text{peak}}, a=1,\cdots,|\mathcal{I'}|$;
  \item Case II: $p_{\pi(a)}^*=P_a^{\text{peak}}, a=1,\cdots,|\mathcal{I'}|-1$, $p_{\pi(\mathcal{I'})}^*$ can be calculated as:
  \[p_{\pi(\mathcal{I'})}^*\!=\!\!\frac{1}{{\alpha}_{\pi(|\mathcal{I'}|)}}\!\!\!\left[\!\!\left(\!\!\frac{\lambda_{\pi(|\mathcal{I'}|)}}{\varepsilon(\theta){\alpha}_{\pi(|\mathcal{I'}|)}
   }\!\!\right)^{\!-\!\frac{1}{\!\varepsilon(\theta)+1}}\!\!\!-\!\!\!\sum\limits_{a=1}^{|\mathcal{I'}|-1\!\!}\!\!\!P_{\pi(a)}^{\text{peak}}{\alpha}_{\pi(a)}\!\!-\!\!1\!\!\right].\]
\end{itemize}

Then, we have to prove that $|\mathcal{I'}|$ is the largest value of $x$ that satisfies the following condition
\begin{equation}
\frac{\lambda_{\pi(x)}}{\varepsilon(\theta){\alpha}_{\pi(x)}}<\left[\sum\limits
_{b=1}^{x-1}P_{\pi(b)}^{\text{peak}}{\alpha}_{\pi(b)}+1\right]^{-{\varepsilon(\theta)
-1}}.
\label{calmaxII}
\end{equation}

First, we show that in both Case I and Case II, for any user $\pi(a)\in \mathcal{I'}$, the above inequality holds. For Case I, as $p_{\pi(|\mathcal{I'}|)}^*=P_{|\mathcal{I'}|}^{\text{peak}}$, by using (\ref{kkt-u}) and (\ref{kkt-delta}), it follows that $\mu_{|\mathcal{I'}|}\geq0$ and $\delta_{|\mathcal{I'}|}=0$. Then, substituting them into (\ref{kkt-p}) yields
\[\begin{array}{l}
\frac{{{\lambda _{\pi (|{I^\prime }|)}}}}{{\varepsilon (\theta ){\alpha _{\pi (|{I^\prime }|)}}}} \le {\left[ {\sum\limits_{b = 1}^{|{I^\prime }|} {P_{\pi (b)}^{{\rm{peak}}}} {\alpha _{\pi (b)}} + 1} \right]^{ - \varepsilon (\theta ) - 1}}\\
 < {\left[ {\sum\limits_{b = 1}^{|{I^\prime }| - 1} {P_{\pi (b)}^{{\rm{peak}}}} {\alpha _{\pi (b)}} + 1} \right]^{ - \varepsilon (\theta ) - 1}}.
\end{array}\]
Hence, the above inequality holds for $a=|\mathcal{I'}|$. From Lemma 1, the left hand side (LHS) of (\ref{calmaxII}) increases as $x$ increases, while the right hand side of (RHS)  (\ref{calmaxII}) decreases as $x$ increases. Hence, the inequality (\ref{calmaxII}) holds for $x=1,\cdots,|\mathcal{I'}|-1$. As a result, for $x=1,\cdots,|\mathcal{I'}|$, the inequality (\ref{calmaxII}) holds.  A similar proof can be extended to Case II, the details of which are omitted.

Finally, we show that for any RRH $\pi(j), j\in \left\{|\mathcal{I'}|+1,\cdots,I\right\}$, (\ref{calmaxII}) does not hold. Due to the fact that  the LHS of (\ref{calmaxII}) increases as $x$ increases while the RHS of  (\ref{calmaxII}) decreases as $x$ increases, it is sufficient to prove that for RRH $\pi(|\mathcal{I'}|+1)$, (\ref{calmaxII}) does not hold. According to (\ref{kkt-u}) and (\ref{kkt-delta}), we have $\delta^*_{\pi(|\mathcal{I'}|+1)}\geq 0$ and $\mu^*_{\pi(|\mathcal{I'}|+1)}=0$. Then, from (\ref{kkt-p}), it follows that
\[\begin{array}{l}
\frac{{{\lambda _{\pi (|{I^\prime }| + 1)}}}}{{\varepsilon (\theta ){\alpha _{\pi (|{I^\prime }| + 1)}}}} \ge {\left[ {\sum\limits_{b = 1}^{|{I^\prime }|} {p_{\pi b}^*} {\alpha _{\pi (b)}} + 1} \right]^{ - \varepsilon (\theta ) - 1}}\\
 \ge {\left[ {\sum\limits_{b = 1}^{|{I^\prime }|} {P_{\pi (b)}^{{\rm{peak}}}} {\alpha _{\pi (b)}} + 1} \right]^{ - \varepsilon (\theta ) - 1}},
\end{array}\]
which does not satisfy inequality (\ref{calmaxII}). Therefore, the proof is completed.\hfill\rule{2.7mm}{2.7mm}

\vspace{-0.4cm}
\section{Proof of Theorem2}\label{theorem2}
 For any given $\tilde{\bm\lambda}$, denote the optimal solution of Problem (\ref{dual-function}) by ${\bf{P}}_{\tilde {\bm\lambda}}^*(\bm\nu) = {\left[ {{ p}_{(1,\tilde {\bm\lambda} )}^*(\bm\nu) , \cdots ,{ p}_{(I,\tilde {\bm\lambda}) }^*(\bm\nu) } \right]^T}$  when $ \bm\lambda  =\tilde {\bm\lambda}$. Then, we have
\begin{align}
g(\tilde{\bm\lambda}) &= \mathop {\min }\limits_{{\bf{P}}\in {\cal P}} \left\{ {\mathbb{E}_{\bm\alpha} }[{{(1 + \sum\limits_{i \in {\cal I}} {{p_i}(\bm\nu)} {{ {\bm\alpha} }_i})}^{ - \varepsilon (\theta )}}] +\right.\nonumber\\
 &\qquad\qquad\left.\sum\limits_{i \in {\cal I}} {\tilde\lambda _i} ({\mathbb{E}_{\bm\alpha }}[{p_i(\bm\nu)}] - P_i^{{\rm{avg}}}) \right\}\label{first}\\
&\le {\mathbb{E}_{\bm\alpha} }[{(1 + \sum\limits_{i \in {\cal I}}{ p}_{(i, \bm\lambda^{(k)} )}^*(\bm\nu){{ {\bm\alpha} }_i})^{ - \varepsilon (\theta )}}] +\nonumber\\
& \qquad\qquad\sum\limits_{i \in {\cal I}} {\tilde\lambda _i} ({\mathbb{E}_{\bm\alpha} }[{ p}_{(i, \bm\lambda^{(k)} )}^*(\bm\nu)] - P_i^{{\rm{avg}}})\label{twoho}\\
&= \!g({\bm\lambda}^{(k)}) \!+\!\! \sum\limits_{i \in {\cal I}} \!\!{\left( {\tilde \lambda_i}\!-\!{\lambda _i^{(k)}} \right)} ({\mathbb{E}_{\bm\alpha} }\![{ p_{(i,\bm\lambda^{(k)})}^*(\bm\nu)}] \!-\! P_i^{{\rm{avg}}})\label{third}\\
&=g({\bm\lambda ^{(k)}})\! +\! {\left( {\tilde{\bm\lambda}\! -\! {\bm\lambda ^{(k)}}} \right)^T}({\mathbb{E}_{\bm\alpha} }[{\bf{P}}_{\bm\lambda^{(k)}}^*(\bm\nu)]\! -\! {\tilde{\bf{P}}^{{\rm{avg}}}}),\label{oijfe}
\end{align}
where (\ref{first}) uses the definition of $g(\tilde{\bm\lambda }) $ in (\ref{dual-function}), (\ref{twoho}) follows since $ {\bf{P}}_{\bm\lambda^{(k)}}^*(\bm\nu)$ is not the optimal solution of Problem (\ref{first}) when $ \bm\lambda  = \tilde{\bm\lambda}$.\hfill\rule{2.7mm}{2.7mm}

\vspace{-0.4cm}
\section{The Average Power for the Case of $I=1$}\label{theorem3}
\begin{figure}
\centering
\includegraphics[width=3.3in]{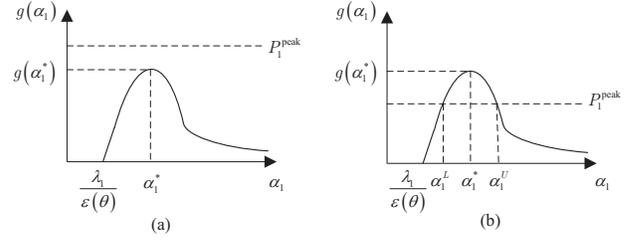}
\vspace{-0.3cm}
\caption{Two cases for computing the average power.}.
\label{casehiohe}
\end{figure}
Let us now define the following function
\begin{equation}\label{dijeo}
g({\alpha _1}) = \frac{1}{{{{\left( {\frac{{{\lambda _1}}}{{\varepsilon (\theta )}}} \right)}^{\frac{1}{{1 + \varepsilon (\theta )}}}}\alpha _1^{\frac{{\varepsilon (\theta )}}{{1 + \varepsilon (\theta )}}}}} - \frac{1}{{{\alpha _1}}}.
\end{equation}
Taking the first-order derivative of $g({\alpha _1})$ with respect to ${\alpha _1}$ and setting it to zero yields:
\begin{equation}\label{jdeo}
  g'({\alpha _1}) =  - \frac{1}{{{{\left( {\frac{{{\lambda _1}}}{{\varepsilon (\theta )}}} \right)}^{\frac{1}{{1 + \varepsilon (\theta )}}}}\alpha _1^{1{\rm{ + }}\frac{{\varepsilon (\theta )}}{{1 + \varepsilon (\theta )}}}}}\frac{{\varepsilon (\theta )}}{{1 + \varepsilon (\theta )}} + \frac{1}{{\alpha _1^2}} = 0.
\end{equation}
By solving (\ref{jdeo}), we can obtain the solution $\alpha _1^{\rm{*}} = \frac{{{\lambda _1}}}{{\varepsilon (\theta )}}{\left( {\frac{{1 + \varepsilon (\theta )}}{{\varepsilon (\theta )}}} \right)^{1 + \varepsilon (\theta )}}$. It can be easily verified that the function $g({\alpha _1})$ first increases in the region of ${\alpha _1} \in \left[ {\frac{{{\lambda _1}}}{{\varepsilon (\theta )}},\alpha _1^{\rm{*}}} \right]$ and then decreases when ${\alpha _1} \in \left[ {\alpha _1^{\rm{*}},\infty } \right]$. Hence, we can obtain the maximum value of the function $g({\alpha _1})$ by substituting $\alpha _1^{\rm{*}}$ into (\ref{dijeo}), which yields
\begin{equation}\label{deafr}
  g(\alpha _1^{\rm{*}}) = \frac{1}{{{\lambda _1}}}{\left( {\frac{{\varepsilon (\theta )}}{{1 + \varepsilon (\theta )}}} \right)^{1 + \varepsilon (\theta )}}.
\end{equation}
It can be readily verified that when ${\alpha _1} \to \infty $, $g({\alpha _1}) \to 0$. In addition, $g\left( {\frac{{{\lambda _1}}}{{\varepsilon (\theta )}}} \right) = 0$. Hence, depending on the comparative value between $g(\alpha _1^{\rm{*}})$ and $P_1^{\rm{peak}}$, two cases should be considered, when computing the average power as illustrated in Fig.~\ref{casehiohe}.  The details are given as follows:
\newcounter{mytempeqncnt1}
\begin{table*}[!t]
\normalsize
\begin{align}\label{deae}
{\mathbb{E}_{{\alpha _1}}}\left\{ {{p_1}} \right\}
 = \left\{ \begin{array}{l}
 \frac{{{{\left( {\frac{{\varepsilon \left( \theta  \right)}}{{{\lambda _1}}}} \right)}^{\frac{1}{{1 + \varepsilon \left( \theta  \right)}}}}{{\left( {\frac{m}{{{{\bar \alpha }_1}}}} \right)}^{\frac{{\varepsilon \left( \theta  \right)}}{{1 + \varepsilon \left( \theta  \right)}}}}}}{{\Gamma \left( m \right)\left( {m - \frac{{\varepsilon \left( \theta  \right)}}{{1 + \varepsilon \left( \theta  \right)}}} \right)}}\left[ { - {{\left( {\frac{{{\lambda _1}m}}{{{{\bar \alpha }_1}\varepsilon \left( \theta  \right)}}} \right)}^{m - \frac{{\varepsilon \left( \theta  \right)}}{{1 + \varepsilon \left( \theta  \right)}}}}{e^{ - \frac{{{\lambda _1}m}}{{{{\bar \alpha }_1}\varepsilon \left( \theta  \right)}}}} + \Gamma \left( {m + \frac{1}{{1 + \varepsilon \left( \theta  \right)}},\frac{{{\lambda _1}m}}{{{{\bar \alpha }_1}\varepsilon \left( \theta  \right)}}} \right)} \right]\\
\qquad - \frac{m}{{(m - 1)\Gamma (m){{\bar \alpha }_1}}}\left[ { - {{\left( {\frac{{{\lambda _1}m}}{{{{\bar \alpha }_1}\varepsilon \left( \theta  \right)}}} \right)}^{m - 1}}{e^{ - \frac{{{\lambda _1}m}}{{{{\bar \alpha }_1}\varepsilon \left( \theta  \right)}}}} + \Gamma \left( {m,\frac{{{\lambda _1}m}}{{{{\bar \alpha }_1}\varepsilon \left( \theta  \right)}}} \right)} \right],{\rm{when}}\  m \ne 1\\
{\left( {\frac{{\varepsilon \left( \theta  \right)}}{{{\lambda _1}}}} \right)^{\frac{1}{{1 + \varepsilon \left( \theta  \right)}}}}\bar \alpha _1^{ - \frac{{\varepsilon \left( \theta  \right)}}{{1 + \varepsilon \left( \theta  \right)}}}\Gamma \left( {\frac{1}{{1 + \varepsilon \left( \theta  \right)}},\frac{{{\lambda _1}}}{{{{\bar \alpha }_1}\varepsilon \left( \theta  \right)}}} \right) + \frac{1}{{{{\bar \alpha }_1}}}{\rm{Ei}}\left( { - \frac{{{\lambda _1}}}{{{{\bar \alpha }_1}\varepsilon \left( \theta  \right)}}} \right),\qquad{\rm{when}}\  m = 1,
\end{array} \right.
\end{align}
\hrulefill
\end{table*}

\emph{Case I: $g(\alpha _1^{\rm{*}})\leq P_1^{\rm{peak}}$.} In this case, the transmit power is always lower than the peak power as shown in Fig.~\ref{casehiohe}. Hence, the peak power constraints are redundant and can be removed. Thus, the average power can be expressed as
\[{\mathbb{E}_{{\alpha _1}}}\left\{ {{p_1}} \right\} = \int_{\frac{{{\lambda _1}}}{{\varepsilon (\theta )}}}^\infty  {g({\alpha _1})f} ({\alpha _1})d{\alpha _1},\]
which can be expanded as (\ref{deae}), where ${\rm{Ei}}(x) \!=\!  -\! \int_{ - x}^\infty  {\left( {{e^{ - t}}/t} \right)dt} $ is the exponential integral function, while $\Gamma \left( {s,x} \right) \!=\! \int_x^\infty  {{t^{s - 1}}{e^{ - t}}} dt$ is the upper incomplete gamma function.

\emph{Case II: $g(\alpha _1^{\rm{*}})>P_1^{\rm{peak}}$.} In this case, there must exist two solutions that satisfy $g({\alpha _1}) = P_1^{{\rm{peak}}}$ as shown in Fig.~\ref{casehiohe}. Denote these two solutions as $\alpha_1^L$ and $\alpha_1^U$ with
$\alpha_1^L<\alpha_1^U$. These two solutions can be numerically obtained by existing algorithms, such as the classic bisection search method. As seen from Fig.~\ref{casehiohe}, the transmit power is equal to $g({\alpha _1})$ when $\alpha _1\in \left[ {\frac{{{\lambda _1}}}{{\varepsilon \left( \theta  \right)}},\alpha _1^L} \right]$ and $\left[ {\alpha _1^U,\infty } \right]$, equal to $P_1^{\rm{peak}}$ when ${\alpha _1} \in \left[ {\alpha _1^L,\alpha _1^U} \right]$. Hence, the average power can be expressed as:
\begin{equation}\label{DEOI}
  \begin{array}{l}
\mathbb{E}{_{{\alpha _1}}}\left\{ {{p_1}} \right\} = \underbrace {\int_{\frac{{{\lambda _1}}}{{\varepsilon (\theta )}}}^{\alpha _1^L} {g({\alpha _1})f} ({\alpha _1})d{\alpha _1}}_{{O_1}} + \underbrace {P_1^{{\rm{peak}}}\int_{\alpha _1^L}^{\alpha _1^U} f ({\alpha _1})d{\alpha _1}}_{{O_2}}\\
\qquad\qquad\quad + \underbrace {\int_{\alpha _1^U}^\infty  {g({\alpha _1})f} ({\alpha _1})d{\alpha _1}}_{{O_3}}.
\end{array}
\end{equation}
Note that $O_1$ and $O_3$ can be derived similarly as the average power in Case I, which are omitted for simplicity. We only provide the expression of $O_2$ as follows:
\[{O_2} = \frac{{P_1^{{\rm{peak}}}}}{{\Gamma (m)}}\left[ {\gamma \left( {m,\frac{m}{{{{\bar \alpha }_1}}}\alpha _1^U} \right) - \gamma \left( {m,\frac{m}{{{{\bar \alpha }_1}}}\alpha _1^L} \right)} \right],\]
where $\gamma \left( {s,x} \right) \!=\! \int_0^x {{t^{s - 1}}{e^{ - t}}} dt$ is the lower incomplete gamma function.

For the case of a single RRH, there is only one dual variable  $\lambda _1$, which can be obtained by solving equation ${\mathbb{E}_{{\alpha _1}}}\left\{ {{p_1}} \right\}=P_1^{\rm{avg}}$ with one-dimension bisection search method.\hfill\rule{2.7mm}{2.7mm}

\vspace{-0.4cm}
\section{Proof of Lemma 3}\label{lemma4}
For $x\geq0$, the second-order derivative of $h(x)$ with respect to (w.r.t.) $x$ is given by $h''(x)=a^2b(b-1)(1+ax)^{b-2}$, which is positive since $a>0,b>1$. Hence, $h(x)$ is a convex function, when $x\geq0$. Combining with the fact that $h(0)=1$, there are only three possible curves for $h(x)$ when $x\geq0$, as shown in Fig.~\ref{caseI1}. Next, we derive the conditions for each case.
\begin{enumerate}
  \item Case 1: In this case, the first-order derivative of $h(x)$ is positive when $x=0$, i.e., $h'(0) \ge 0$, which yields the condition for Case 1 in Lemma 3;
  \item Case 2: The first-order derivative of $h(x)$ is negative for $x=0$, i.e., $h'(0)< 0$, which yields the condition of $ab-c<0$. The minimum point of $h(x)$ is achieved by solving equation $h'(x)=0$, and the solution is given by $x^* = \frac{1}{a}\left( {{{\left( {\frac{c}{{ab}}} \right)}^{\frac{1}{{b - 1}}}} - 1} \right)$. Then, according to Fig.~\ref{caseI1}-b, $h(x^*) \ge 0$ should hold, which leads to the second part of the conditions in Case 2 in Lemma 3;
  \item Case 3: The proof is similar to that in Case 2, and is omitted for simplicity.\hfill\rule{2.7mm}{2.7mm}
\end{enumerate}

\vspace{-0.4cm}
\section{Proof of $\frac{{{\lambda _1}}}{{\varepsilon \left( \theta  \right)}}<\alpha _1^l$}\label{daxiaorelation}
It is easy to verify that $h\left( {{{{\lambda _1}} \mathord{\left/
 {\vphantom {{{\lambda _1}} {\varepsilon \left( \theta  \right)}}} \right.
 \kern-\nulldelimiterspace} {\varepsilon \left( \theta  \right)}}} \right) > 0$ when $a, b, c$ are given in (\ref{abc}). Hence, when the third condition of Lemma 3 given in Condition C1 is satisfied, according to Fig.~\ref{caseI1}-c, ${\lambda _1}/\varepsilon \left( \theta  \right)$ must fall into two regions: 1) $0 < {\lambda _1}/\varepsilon \left( \theta  \right) < \alpha _1^l$; 2) ${\lambda _1}/\varepsilon \left( \theta  \right) > \alpha _1^u$. In the following, we prove that the probability of ${\lambda _1}/\varepsilon \left( \theta  \right)$ falling into the second region is zero by using the method of contradiction.

 Suppose that  ${\lambda _1}/\varepsilon \left( \theta  \right) > \alpha _1^u$ holds. Then, according to Fig.~\ref{caseI1}-c, ${\lambda _1}/\varepsilon \left( \theta  \right)$ must be larger than the optimal point, i.e., ${\lambda _1}/\varepsilon \left( \theta  \right)>x^*$, where $x^*$ is given in Appendix \ref{lemma4}. By substituting $a,b,c$ in (\ref{abc}) into this inequality and after some further simplifications, we have
 \begin{equation}\label{dejoj}
   {\left( {\frac{{P_1^{{\rm{peak}}}{\lambda _1}}}{{\varepsilon \left( \theta  \right)}} + 1} \right)^{\varepsilon \left( \theta  \right)}} > \frac{{\varepsilon \left( \theta  \right)}}{{P_1^{{\rm{peak}}}{\lambda _1}\left( {1 + \varepsilon \left( \theta  \right)} \right)}}.
 \end{equation}
Additionally, by inserting $a,b,c$ in (\ref{abc}) into Condition C1 and after some further simplifications,  we have
\begin{equation}\label{freA}
P_1^{{\rm{peak}}}{\lambda _1}< {\left( {\frac{{\varepsilon \left( \theta  \right)}}{{1 + \varepsilon \left( \theta  \right)}}} \right)^{1 + \varepsilon \left( \theta  \right)}} < 1.
\end{equation}
By substituting (\ref{freA}) into the right hand side of (\ref{dejoj}) and with some simple further operations, one obtains ${\lambda _1}P_1^{{\rm{peak}}}>1$, which contradicts with (\ref{freA}) that ${\lambda _1}P_1^{{\rm{peak}}} < 1$. Hence, the assumption that ${\lambda _1}/\varepsilon \left( \theta  \right) > \alpha _1^u$ cannot hold, which completes the proof.\hfill\rule{2.7mm}{2.7mm}

\vspace{-0.4cm}
\section{Derivations of $T_{{\rm{RRH}}_1}^{{\rm{C2}}}$ when $m$ is a positive integer}\label{derivaitons}
The closed-form expression of $T_{{\rm{RRH}}_1}^{{\rm{C2}}}$ for the case of $m=1$  can be easily obtained by inserting $m=1$ into (\ref{conditw}). In the following, we only focus on the case when $m\geq 2$.

When $m\geq2$, function $\gamma \left( {m,x} \right)$ can be expanded as [8.352.1, Page 899, \cite{gradshteyn2014table}]
\begin{equation}\label{lowerincom}
\gamma \left( {m,x} \right) = (m - 1)!\left( {1 - {e^{ - x}}\sum\limits_{l = 0}^{m - 1} {\frac{{{x^l}}}{{l!}}} } \right).
\end{equation}
By plugging the above expression with $x=W\alpha _1$ into (\ref{conditw}), we obtain the average power for RRH 1 as follows:
\begin{align}
T_{{\rm{RRH}}_1}^{{\rm{C2}}}& \!= \!\!\underbrace {\frac{Z}{{{U^V}}}\!\!\int_U^\infty\!\! \! {\alpha _1^{V + m - 2}{e^{ - \frac{m}{{{{\bar \alpha }_1}}}{\alpha _1}}}\!d{\alpha _1}} }_{{J_1}}\!-\!\! \underbrace {Z\!\!\!\int_U^\infty\!\!  {\alpha _1^{m - 2}{e^{ - \frac{m}{{{{\bar \alpha }_1}}}{\alpha _1}}}d{\alpha _1}} }_{{J_4}}\nonumber\\
 &- \underbrace {\frac{Z}{{{U^V}}}\sum\limits_{l = 0}^{m - 1} {\frac{{{W^l}}}{{l!}}\int_U^\infty  {\alpha _1^{l + V + m - 2}{e^{ - Y{\alpha _1}}}d{\alpha _1}} } }_{{J_3}}\nonumber\\
 &+ \underbrace {Z\sum\limits_{l = 0}^{m - 1} {\frac{{{W^l}}}{{l!}}\int_U^\infty  {\alpha _1^{l + m - 2}{e^{ - Y{\alpha _1}}}d{\alpha _1}} } }_{{J_2}} .\nonumber
\end{align}
With some simple variable substitutions, the closed-form expressions of $J_1,J_2,J_3$ and $J_4$ can be easily calculated as in (\ref{J1J2}) and (\ref{J3J4}), respectively.\hfill\rule{2.7mm}{2.7mm}
\end{appendices}

\
\






\vspace{-0.6cm}
\bibliographystyle{IEEEtran}
\bibliography{myre}


\end{document}